\documentclass[%
 aip,
 amsmath,amssymb,amsfonts,
 reprint,%
]{revtex4-1}

\usepackage{graphicx}
\usepackage{dcolumn}
\usepackage{bm}
\usepackage{hyperref}

\usepackage[utf8]{inputenc}
\usepackage[T1]{fontenc}
\usepackage{mathptmx}
\usepackage{etoolbox}
\usepackage{amsmath,amsthm}
\usepackage{cleveref}
\usepackage{booktabs}
\newtheorem{theorem}{Theorem}

\usepackage{xcolor}

\makeatletter
\def\@email#1#2{%
 \endgroup
 \patchcmd{\titleblock@produce}
  {\frontmatter@RRAPformat}
  {\frontmatter@RRAPformat{\produce@RRAP{*#1\href{mailto:#2}{#2}}}\frontmatter@RRAPformat}
  {}{}
}%
\makeatother

\usepackage{tikz}
\usetikzlibrary{shapes.geometric, arrows}

\tikzstyle{startstop} = [rectangle, rounded corners, 
minimum width=3cm, 
minimum height=1cm,
text centered, 
draw=black, 
fill=red!30]

\tikzstyle{io} = [trapezium, 
trapezium stretches=true, 
trapezium left angle=70, 
trapezium right angle=110, 
minimum width=3cm, 
minimum height=1cm, text centered, 
draw=black, fill=blue!30]

\tikzstyle{process} = [rectangle, 
minimum width=3cm, 
minimum height=1cm, 
text centered, 
text width=3cm, 
draw=black, 
fill=orange!30]

\tikzstyle{decision} = [diamond, 
minimum width=3cm, 
minimum height=1cm, 
text centered, 
draw=black, 
fill=green!30]
\tikzstyle{arrow} = [thick,->,>=stealth]

\begin{document}

\preprint{AIP/123-QED}

\title{Transport Impacts of Resonant Island Chains in Fusion Plasmas}
\author{Sidney D. V. Williams}
\affiliation{Center for Energy Research, University of California San Diego, La Jolla, CA, 92093}
\email{siw038@ucsd.edu}
\author{Kevin Mitchell} 
\author{Ethan Custodio}
\affiliation{University of California Merced, Merced, CA, USA}
\author{Dmitri M. Orlov}
\affiliation{Center for Energy Research, University of California San Diego, La Jolla, CA, 92093}

\date{\today}

\begin{abstract}
Magnetic field-line transport in magnetically confined plasmas is commonly modeled as diffusion through stochastic magnetic layers generated by resonant perturbations. However, experimentally relevant magnetic configurations frequently exhibit transport rates that are substantially slower than predicted by fully stochastic models, indicating the presence of unresolved dynamical barriers. In this work, we investigate the role of nested resonant island chains and their associated invariant manifolds in regulating transport within the edge magnetic topology of DIII-D tokamak discharge \#171491 subjected to strong $n=3$ resonant magnetic perturbations. Using field-line tracing, Poincar\'e analysis, and manifold calculations, we identify a hierarchy of homoclinic and heteroclinic tangles associated with both period-one and period-ten hyperbolic points. A systematic construction based on primary intersection points is developed to extract partial transport barriers from the full manifold geometry and divide the chaotic region into dynamically distinct domains. Monte Carlo simulations reveal that field-line escape exhibits a bi-exponential character when only the outermost period-one boundary is considered, reflecting the presence of a long-lived "sticky" region associated with the period-ten island chain. When the inner period-ten boundary is explicitly incorporated into the Monte Carlo, the escape process becomes approximately single exponential and is accurately described by turnstile lobe transport. The measured escape rate agrees with predictions obtained from the turnstile lobe area and the symplectic structure of the field-line map. These results demonstrate that nested tangles provide a quantitative transport skeleton governing magnetic-field-line escape and establish a framework for analyzing transport barriers, cantori, and lobe dynamics in realistic fusion-plasma magnetic configurations.
\end{abstract}

\maketitle

%

\begin{quotation}
In the search for efficient nuclear fusion, tokamaks are among 
the most extensively studied approaches to magnetic confinement. Their performance depends critically on controlling how heat and particles move from the hot plasma core toward material surfaces. Three-dimensional magnetic perturbations, while useful for controlling plasma instabilities, can create chaotic magnetic-field regions that enhance this transport. Here we use ideas from nonlinear dynamics—particularly invariant manifolds, turnstile lobes, and phase-space transport—to identify barriers embedded within this chaos. These structures divide the magnetic field into regions with distinct transport properties and provide a geometric way to predict field-line escape rates. The results show that even strongly chaotic magnetic configurations can retain organized structures that substantially regulate transport.


\end{quotation}

\section{Introduction}

Although contemporary fusion research is often driven by practical engineering goals, fusion and plasma physics have also produced foundational equations and tools for the study of chaos \cite{boozer2005physics, meiss1992symplectic, kuramoto1978diffusion, sivashinsky1977nonlinear}. Perhaps the most relevant example is Chirikov's standard map 
\begin{equation}\label{standard}
\begin{split}
    I_{n+1}=I_n+K\sin\theta_n,\mod(2\pi)\\
    \theta_{n+1}=I_n+\theta_n+K\sin\theta_n,\mod(2\pi)
\end{split}
\end{equation}
which was introduced to describe chaotic charged-particle dynamics in magnetic mirror traps---open-ended plasma-confinement devices in which stronger magnetic fields near the ends reflect particles toward the center \cite{chirikov1979universal}.

Despite its simple form, the Standard Map captures the essential physics of non-integrability generic in fusion plasmas. When $K=0$, the system becomes integrable, leaving the "momentum" $I$ as an invariant, and $\theta$ rotating around the $I$-radius circle. This situation is exactly analogous to the toroidal axisymmetry assumption in tokamak fusion reactors without purposeful 3D perturbations, where magnetic field lines follow integrable trajectories and lie on unbroken nested surfaces \cite{boozer2005physics, abdullaev2006construction}. These so-called \emph{flux surfaces} have constant poloidal magnetic flux $\psi_p$, defined as the flux of the magnetic field through a horizontal disk, centered at the tokamak's symmetry axis (see fig.\ref{fig:equilpoincoord})\cite{hazeltine2003plasma}. The value of this flux at the plasma's magnetic axis (the central ring of the nested flux surfaces) is subtracted off to define an effective radial coordinate $\psi_N$. 

In modern tokamaks, externally applied three-dimensional magnetic perturbations are used to mitigate or suppress edge-localized modes (ELMs), thereby reducing the intense transient heat loads they impose on plasma-facing components \cite{evans2004suppression}.  When these 3D field perturbations are imposed, magnetic field line trajectories cease to be integrable, and much like the trajectories of the Standard Map, flux surfaces distort and break according to Kolmogorov-Arnold-Moser (KAM) theory. Again, like the Standard Map, resonant perturbations induce island structure, and overlapping resonances produce a chaotic sea of magnetic field lines which strongly affects overall plasma dynamics within the fusion device. 

Magnetic field line transport through the mixed phase space generated by intermixing of stable islands and chaotic field lines is central to an effective utilization of 3D fields in magnetic confinement fusion reactors. Additionally, it provides a laboratory-realizable test bed of dynamical systems and chaos in mixed phase spaces, which---as will be shown in the subsequent section---is represented by a quite small and simple set of nonlinear ordinary differential equations.

In this study, we begin to describe transport through the mixed phase space utilizing the topology of heteroclinic tangles. Tangles have been studied for both purely mathematical maps, without direct physical analogs, and maps derived from physical applications \cite{Mitchell2003, Mitchell2004, Jakubowski2007, Wingen2009Footprints, Smiet2025}.  For example, a recent work used tangles to quantify ionization in chaotic hydrogen, in which the tangle was used to compute thousands of periodic orbits that were used for analysis\cite{custodio2025hydrogen}. 

Here, we do an introductory analysis to verify the feasibility of these methods and acquire initial results.  We use the tangles to define regions of phase space that we will measure transport between with both Monte Carlo methods and lobe area estimates. We quantify escape from and between those regions. This marks a novel step forward in the analysis of chaotic transport of plasma in tokamak reactors and the basis for a further detailed study of the dynamics, which will involve partitioning of the phase space and discovery of periodic orbits. 

This work is organized as follows: in section II, the Hamiltonian structure of magnetic confinement fusion reactor magnetic field lines is elucidated. In section III, DIII-D shot \#171491 is explored; this includes finding the period-10 orbit associated with the outermost magnetic island chain and calculating the stable and unstable manifolds of the chain, along with the stable and unstable manifolds associated with the period-1 primary x-point (see Fig. \ref{fig:equilpoincoord}), which defines the magnetic equilibrium. Then, the stable and unstable manifolds are systematically reduced to dynamical barriers which are used in Monte Carlo simulations of magnetic field line escape rate. The found escape rates are then related to a simple lobe dynamics estimate, yielding an intuitive geometrical picture of the dynamics of chaotic field lines. Section IV concludes the paper by summarizing the effort, connecting the results and interpretation to outstanding problems in the field of fusion plasma physics, and suggesting future lines of investigation. Appendix A clearly lays out the algorithm used to reduce the full stable and unstable manifold structures through utilization of primary intersection points (PIPs). Finally, Appendix B thoroughly describes the statistical framework used to validate the choice of single or double exponential escape rates, and the resulting fits.
\section{Hamiltonian Chaos in Fusion Reactors}
Central to the standard paradigm of confinement in tokamaks is the inherent Hamiltonian structure in the magnetic field. Without any toroidal variation in the magnetic structure, the magnetic field lines, which impose cyclotron motion upon the charged particles \cite{goldston1995introduction} that make up the plasma, follow simple, integrable paths on nested flux surfaces. However, when integrability is damaged, the nested flux surfaces break, and the natural transport barriers degrade. This provides an excellent physical testbed on which dynamical systems theory may be applied and experimentally diagnosed. Additionally, the tools from dynamical systems may be utilized to rigorously understand why plasma confinement may be maintained even without fully intact flux surfaces.

\subsection{Hamiltonian Mechanics of Magnetic Field Lines}

\begin{figure}[t]\centering
    \includegraphics[width=0.7\columnwidth]{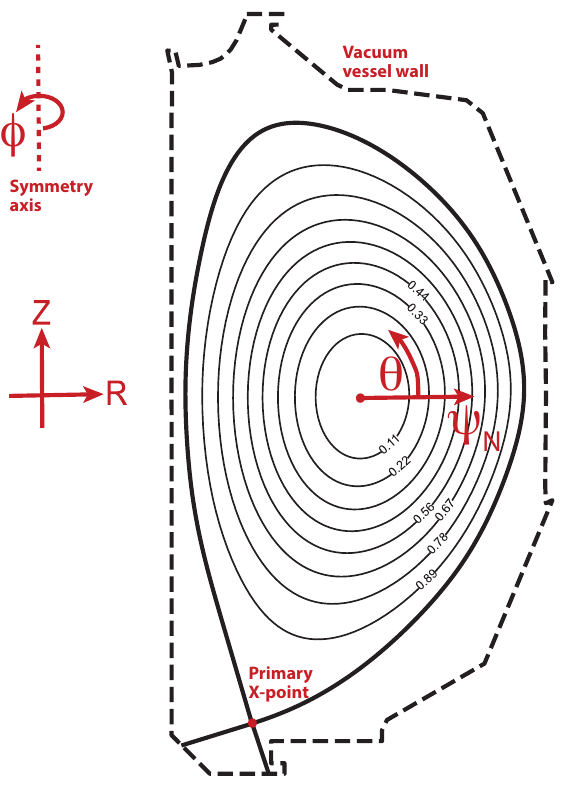}
    \caption{Tokamak geometry and coordinate conventions used in this work. The main panel shows a poloidal cross-section at fixed toroidal angle $\phi$. The cylindrical coordinates $R$ and $Z$ locate a point geometrically, whereas the normalized poloidal flux $\psi_N=\frac{\psi_p-\psi_{axis}}{\psi_{sep}-\psi_{axis}}$ labels the nested equilibrium magnetic-flux surfaces and $\theta$ measures poloidal position around a surface. The separatrix forms the boundary of the closed-flux region and passes through the primary X-point. The dashed outline denotes the vacuum-vessel wall. The three-dimensional tokamak geometry is obtained by revolving the poloidal cross-section around the central symmetry axis.}
    \label{fig:equilpoincoord}
\end{figure}

The trajectory $\mathbf{x}(\lambda)$ of a magnetic field line satisfies the variational principle 
\cite{cary1983noncanonical}
\begin{equation}\label{action}
    \delta\int\mathbf{A}(\mathbf{x})\cdot \frac{d\mathbf{x}}{d \lambda} \; d\lambda=0,
\end{equation}
where $\mathbf{A}$ is the vector potential.  The above equation is equivalent to the constraint $\mathbf{B} \times d\mathbf{x}/d \lambda = 0$, i.e. the tangent to the trajectory is parallel to the magnetic field.  To cast this in Hamiltonian form, we write
\begin{equation}\label{action}
    \int\mathbf{A}(\mathbf{x})\cdot d\mathbf{x}=\int p dq-Hd\tau,
\end{equation}
where we must identify the canonical coordinates $p$ and $q$ as well as the Hamiltonian $H$ and "time" $\tau$.  We work in cylindrical coordinates $(R,\phi,Z)$ where $\phi$ is the toroidal angle around the tokamak, $Z$ is the vertical coordinate measured from the midplane, and $R$ is the distance from the central symmetry axis (Fig.~\ref{fig:equilpoincoord}).  Using the gauge freedom of $\mathbf{A}$ to eliminate its $R$ component, we write
\begin{equation}\label{legendre}
    A_Z(R,\phi,Z)dZ+A_{\phi}(R,\phi,Z)d\phi=pdq-Hd\tau.
\end{equation}
Note that the left hand side of eqn \eqref{legendre} contains the vector potential 1-form, and so the components do not have the physical space metric factors (eg. the 1-form component is $A_{\phi}$ not $RA_{\phi}$). Matching components, we identify the canonical position, canonical momentum, Hamiltonian, and "time" as
\begin{equation}\label{can_vars}
    \begin{split}
        q &= Z,\\
        p &= A_Z,\\
        H &= -A_{\phi}, \\
        \tau &= \phi.
    \end{split}
\end{equation}
Working in $RZ$-coordinates, the tangency condition implies
\begin{equation}\label{field_line_eqns}
    \frac{dR}{d\phi}=\frac{RB_R}{B_{\phi}},\quad \frac{dZ}{d\phi}=\frac{RB_Z}{B_{\phi}}.
\end{equation}
These equations can be interpreted as Hamilton's equations in the (noncanonical) $RZ$-coordinates.  As such, they preserve the canonical 2-form $dq \wedge dp$~\cite{arnold1989mathematical}, which written in noncanonical coordinates and restricted to a surface of constant $\phi$ yields,
\begin{equation} \label{poincare-cartan}
    \omega=dq \wedge dp = B_{\phi}dR\wedge dZ.
\end{equation}
This conserved 2-form acts as an area element of phase space. It will be used later to calculate the magnetic field line escape rate through lobe dynamics.

When the magnetic field is toroidally axisymmetric, the magnetic field line trajectories can be described using canonical action-angle variables $(J(\psi_p), \theta)$, where $\psi_p$ is the conserved poloidal flux defined in the Introduction~\cite{hazeltine2003plasma}. The Hamiltonian $H_0(\psi_p)$ defines a set of nested invariant tori with the associated winding number defined as $q(\psi_p) = d\phi/d\theta$. However, as stated in the introduction, many configurations for modern tokamak operation require non-axisymmetric magnetic fields generated via purposeful application of external perturbations:
\begin{align}
    H(\psi_p, \theta, \phi) &= H_0(\psi_p) + \sum_{m,n} \epsilon_{m,n}(\psi_p) \cos(m\theta - n\phi + \chi_{m,n})\\
    \mathbf{B} &=\mathbf{B}_0(\psi_p) + \delta\mathbf{B}
    \label{r1}
\end{align}
As in the example of the Standard Map, these perturbations severely damage the nested flux surface structure. In fact, surfaces that have a rational winding number which resonates with an imposed perturbation suffer magnetic reconnection-driven island formation \cite{abdullaev2006construction, fitzpatrick1991interaction}, an event where resonating field lines are sheared apart and reconnected as a stable island. When many surfaces resonate, or islands get particularly large, island overlap occurs, and a region of chaotic field line trajectories form \cite{lichtenberg2013}. Importantly, the resulting chaotic dynamics are not structureless; invariant structures embedded within the chaotic phase space can provide an underlying skeleton that organizes the dynamics and associated transport~\cite{Williams2025}.

At this point, the realm of classical Hamiltonian chaos has been entered. The physical fusion reactor has effectively generated a classical twistmap with experimentally observable implications to dynamics and transport \cite{evans2007separatrix}. Of particular interest are lobe-like heat-flux patterns observed experimentally in magnetically confined plasmas and magnetic footprint structures associated with invariant manifolds in DIII-D \cite{Jakubowski2007, Wingen2009Footprints}. These structures provide channels through which heat and particles can reach plasma-facing surfaces. This can aid in efficiently spreading heat or cause severe damage depending on the transport properties of the chaotic system formed by the magnetic field lines.

\subsection{The DIII-D Tokamak and the Analyzed Experiment}
The utility of the framework developed in the previous section is that the field-line equations, Eq.~\eqref{field_line_eqns}, are not restricted to a particular idealized model and can be applied directly to magnetic fields from an actual fusion experiment. Here, we apply this framework to an experiment conducted on the DIII-D tokamak in San Diego, California. DIII-D has operated as a major magnetic-confinement fusion research facility since 1986 \cite{Luxon2002} and has accumulated an extensive experimental database comprising more than 200,000 plasma discharges, or ``shots.'' Each shot typically lasts several seconds and is performed as part of a broader experimental campaign designed to investigate specific aspects of plasma behavior.

\begin{figure*}[ht]
\includegraphics[width=2\columnwidth]{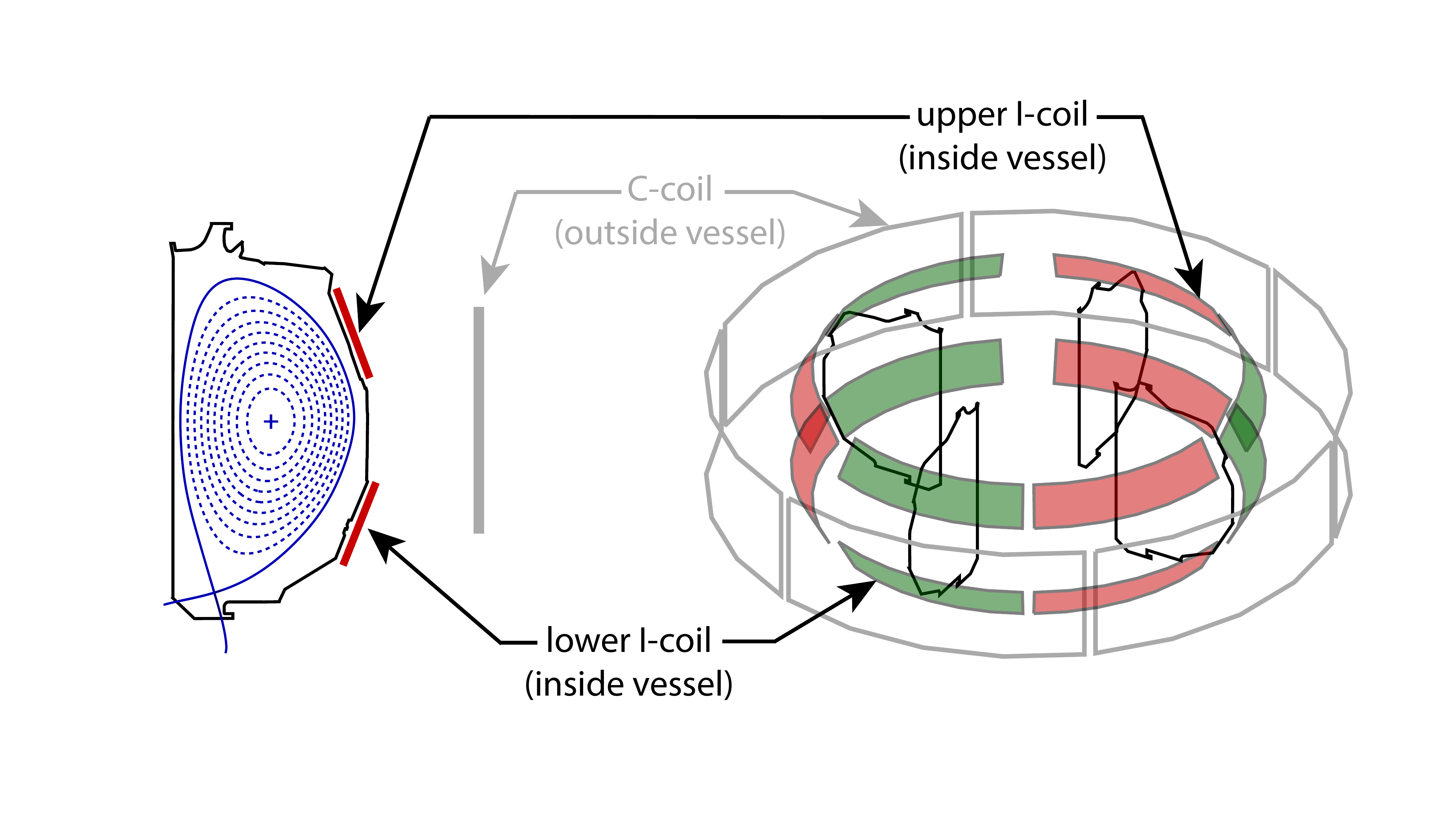}
\caption{\label{fig:DIIID_Coils} Schematic of the auxiliary coil system in the DIII-D tokamak. In this work, only the current produced by the I-coils is considered, but in typical experiments additional coils, such as the grayed-out C-coils, may also contribute.}
\end{figure*}

To investigate the dynamics of magnetic field lines in an experimentally relevant tokamak configuration, and in particular the influence of layered magnetic-island chains on transport, we consider DIII-D discharge \#171491 at $t=3590$~ms. In this experiment, the non-axisymmetric magnetic perturbation is produced by the DIII-D internal control coils (I-coils) \cite{Jackson2003}, shown in Fig.~\ref{fig:DIIID_Coils}. These I-coils are configured to generate an $n=3$ current perturbation, introducing a three-fold toroidal periodicity into the otherwise axisymmetric equilibrium magnetic field via the Biot-Savart law. In the "vacuum approximation," these calculated magnetic perturbations are simply superimposed on top of the axisymmetric equilibrium magnetic field, neglecting any self-consistent plasma response. The conditions this approximation implies are stated at the end of the next section. To summarize, the magnetic field we analyze is based on Eq.~(\ref{r1}), where $\mathbf{B}_0$ is the measured equilibrium field of the plasma, which is axisymmetric and hence integrable.  The perturbation $\delta \mathbf{B}$ is computed from the vacuum Biot-Savart law applied to the geometry of the currents in the I-coils.  This perturbation has three-fold symmetry ($n = 3$.)   

This discharge and time slice are particularly well suited to the present analysis because of the relatively strong applied current perturbation and weak plasma rotation (average toroidal and poloidal motion of the plasma as a whole)\cite{schmitz2019lh}. The strong perturbation produces a well-developed chaotic edge region containing multiple resonant island chains, while the weak rotation reduces plasma screening of the externally applied field. These conditions make the vacuum-field approximation employed here more appropriate and provide an experimentally relevant magnetic configuration sufficiently far from the integrable limit to exhibit the phase-space structures of interest. 

\subsection{Magnetic Islands, Sticky Regions, and MAFOT} 
To explore the chaotic topology generated by controlled nonaxisymmetric perturbations, eqn \eqref{field_line_eqns} is integrated using the MAFOT code \cite{wingen2009highresolution}. MAFOT takes a distribution of initial field line positions and solves Hamilton's equations to track field line positions in cylindrical coordinates $(R,\phi,Z)$, where $\phi$ is the toroidal angle. This framework is extremely flexible, allowing for Monte Carlo simulations of field line trajectories, Poincar\'e section calculation, fixed point determination, and manifold generation. 

\begin{figure}[t]
    \centering
    \includegraphics[width=\columnwidth]{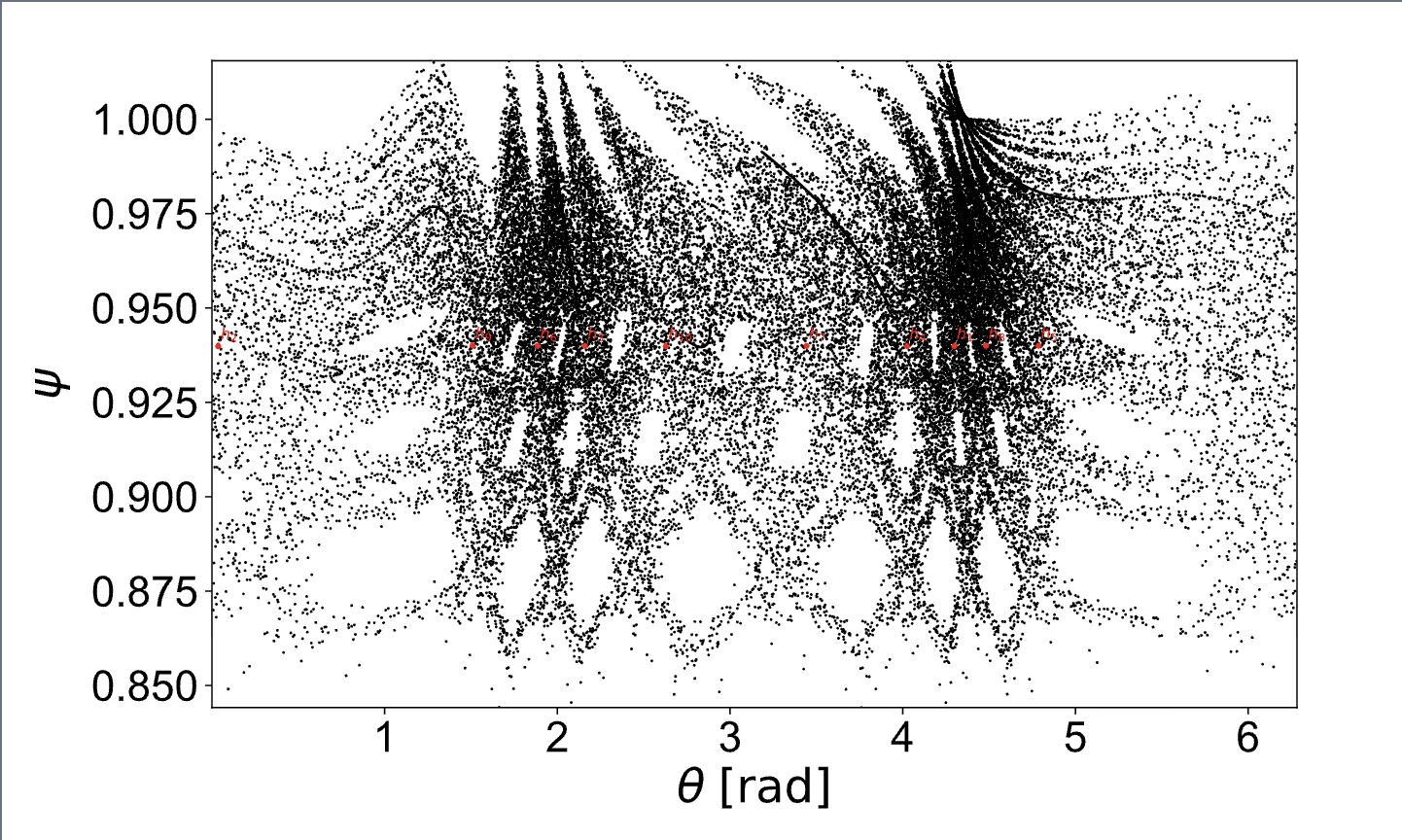}
    \caption{Poincar\'e section of magnetic-field-line trajectories for DIII-D discharge \#171491 at \(t=3590~\mathrm{ms}\), computed with the \(n=3\) I-coil perturbation in the vacuum-field approximation. The black points reveal the layered resonant-island structure near the plasma edge, while the red markers identify the period-10 hyperbolic orbit associated with the outermost intact island chain. This orbit supplies the anchor points used to construct the period-10 resonance-zone boundary.}
    \label{fig:poincare}
\end{figure}

The specific characteristics of chaotic field line transport within tokamaks are strongly dependent on the internal structure of the chaotic magnetic field. In particular, resonant island chains can substantially modify particle transport, with the resulting transport regime depending on the island topology and the location of particles relative to the island O- and X-points \cite{Kostadinova2023, Eskew2026, Khavin2026}. (An O-point is a neutrally stable periodic point and an X-point is a hyperbolic periodic point.)  This structure is largely manifested through stable island formation. As the edge of the outer period-1 resonance zone is approached from the center of the tokamak, more and more flux surfaces resonate with the applied perturbation and are allowed to reconnect. This yields a layered structure of island chains (see Fig. \ref{fig:poincare}).  Interleaved between the islands are X-points whose stable and unstable manifolds mediate transport around the islands. As in dynamical systems, such as the Standard Map, produced from reduced toy models of physics, layered island chains can reduce escape rate, even generating regions of anomalous or "sticky" dynamics \cite{Kostadinova2023} where hyperbolicity breaks down, and an algebraic escape rate is established \cite{sattari2017periodic, meiss1992symplectic}. The extent to which escape is affected is extremely important to understand, as it directly determines heat and particle loss handling in fusion plasmas.

MAFOT also allows for the use of different magnetic field perturbation models. In an ideal (zero resistivity) plasma, when an external current perturbation (which leads to a magnetic perturbation through Maxwell's equations) is applied, electrons generate currents that exactly screen out the perturbation \cite{fitzpatrick1991interaction}, preventing any loss of integrability. However, realistic plasmas have finite resistivity and only engage in partial screening, leading to complex interactions which dictate the resulting structure of the perturbations felt in the plasma \cite{ferraro2012twofluid}. Fortunately, these structures are still just small nonaxisymmetric perturbations that interact with the theory of Hamiltonian chaos through KAM theory in the same way as any other symmetry-breaking perturbation. So, in order to simplify the analysis, we will exclusively consider the "vacuum approximation", which assumes that no screening of the I-coil currents takes place. This approximation is equivalent to assuming slow plasma rotation and/or particularly high resistivity \cite{fitzpatrick1998bifurcated, kim2015ideal}.

\begin{figure}[t]
    \centering
    \includegraphics[width=\columnwidth]{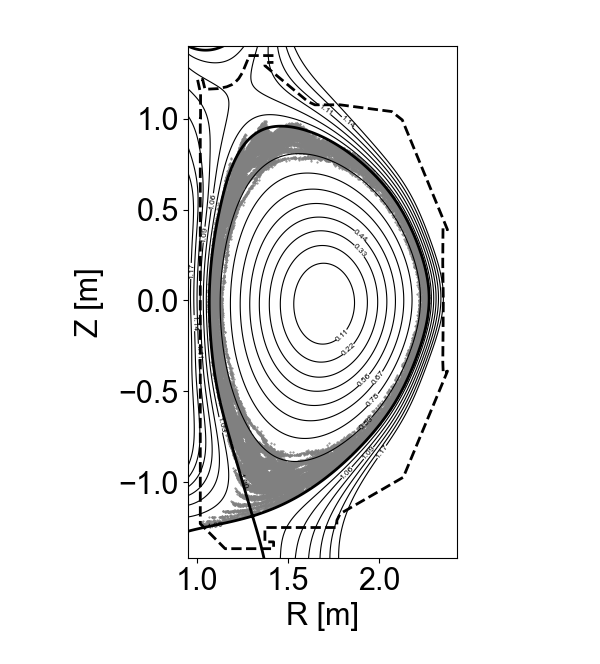}
    \caption{Poloidal cross-section at \(\phi=0\) for DIII-D discharge \#171491 at \(t=3590~\mathrm{ms}\). Thin black curves show contours of the axisymmetric equilibrium flux, the thick black curve denotes the unperturbed separatrix \((\psi_N=1)\), gray points show intersections of perturbed magnetic field lines forming the stochastic edge layer, and the dashed curve marks the vacuum-vessel wall.}
    \label{fig:equilpoin}
\end{figure}

\begin{figure}[t]
\includegraphics[width=\columnwidth,
trim=6cm 0.75cm 6cm 0cm,
clip]{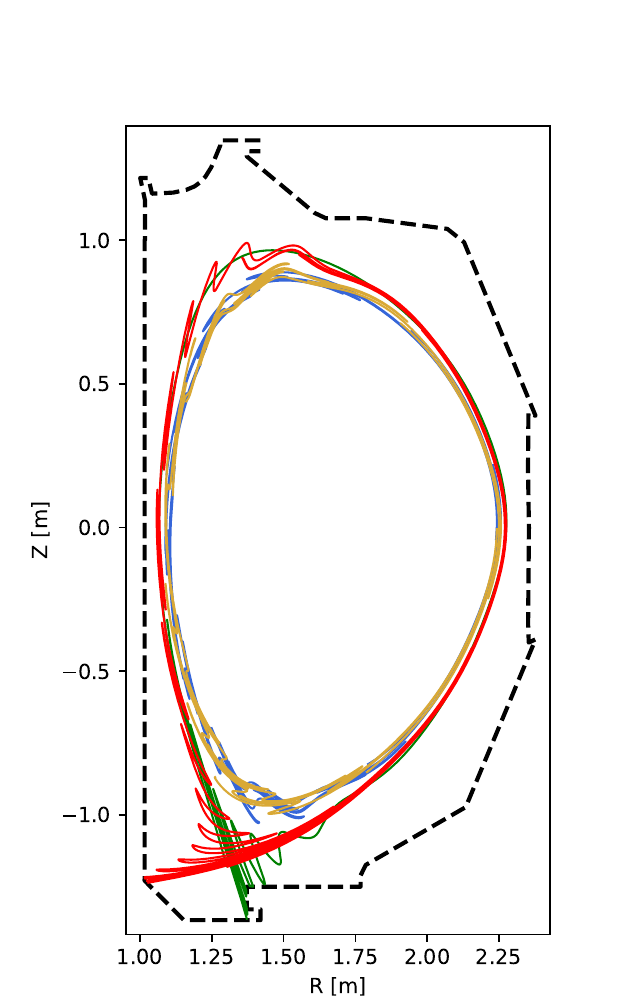}
\caption{\label{fig:fullmani} The full magnetic manifold structure of both the outer period-1 island, and the inner period-10 island chain. The lobes from the outer homoclinic tangle intersect with the walls (dashed line) of the DIII-D tokamak, allowing additional channels through which heat and particles may escape.}
\end{figure}

\begin{figure}[t]
\includegraphics[width=\columnwidth,
trim=6cm 0.75cm 6cm 0cm,
clip]{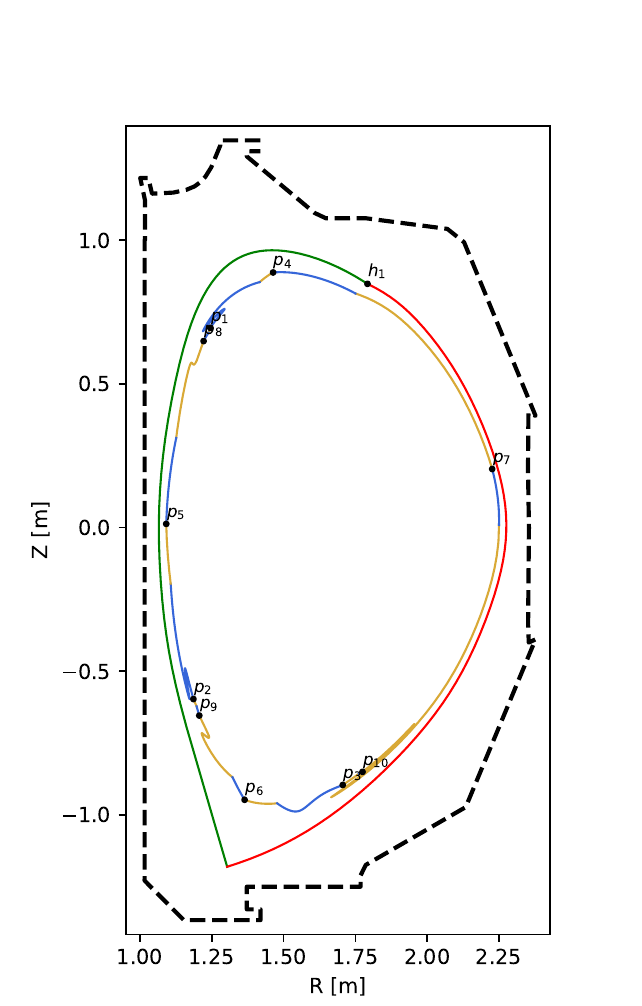}
\caption{\label{fig:stitchedmani} The boundary defined by stitching both the outer period-1 manifolds and the inner period-10 manifolds. The merge points are defined by the pips calculated with the algorithm described in Appendix A. The dashed line is the DIII-D wall.}
\end{figure}

\subsection{Heteroclinic Tangles}\label{heteroclinic_tangles}
Heteroclinic tangles are defined using the anchor points from which they are generated. Anchor points form hyperbolic periodic orbits in the phase space. Two types of invariant manifolds are attached to each anchor point $z_i$, stable and unstable manifolds. All points on the stable manifold converge upon the periodic orbit under forward iteration. Conversely, points on the unstable manifold converge upon the periodic orbit under backward iteration. A point on the stable manifold will forever remain on the stable manifold, mapping infinitely closer to the anchor point. The number of iterations to return to the same manifold is equivalent to the period $n$ of the anchor point. Invariant manifolds are infinitely long and together form a complicated structure called a heteroclinic tangle. The tangle describes the phase space structure and transport within phase space \cite{mackay1987resonances, wiggins1990geometry, romkedar1994homoclinic}. Of key importance are the intersections between the manifolds. By construction, two manifolds of the same stability will never intersect, but two manifolds of different stability will intersect infinitely if they intersect at least once. Classifying intersections is the first step to constructing a complete partition of the phase space \cite{mitchell2012partitioning} and doing a lobe area analysis \cite{mackay1987resonances, romkedar1994homoclinic}. We start by computing a set of intersection points, termed primary intersection points (pips)\cite{wiggins1990geometry, lomeli2000heteroclinic}, from which we construct boundaries in the phase space. 

Starting with a full infinite-length tangle $W$, where $W^U$ and $W^S$ are the unstable and stable manifolds, respectively. $W^U\left(z_i, q_i\right)$ is the open interval of $W^U$ between the two points $z_i, q_i \in W^U$. An intersection is a pip if there are no intersections between the stable and unstable manifolds from $z_i$ to $q_i$. That is, 

$$W^U\left(z_i, q_i\right) \cap W^S\left(z_i, q_i\right) = \emptyset .$$

In order to construct a well-defined escape boundary for a given tangle, it is necessary to compute an iterate cycle of pips so that $\mathcal{M}\left(q_i\right) = q_{i+1}$. Therefore, the manifolds also map to each other around the resonance zone. Additionally, $q_0$ must be a 'strong pip' by satisfying the following condition.

$$ \left[ \bigcup_{i=0}^{n-1} W^U \left( z_i, q_i \right) \right] \cap \left[ \bigcup_{i=0}^{n-1} W^S \left( z_i', q_i \right) \right] = \emptyset .$$

A detailed discussion of how the pips were computed in this paper is included in \hyperref[pip_algo]{Appendix A}. It is worth noting that setting up the system in this way is also essential to constructing a full partition, which can eventually be used to extract periodic orbits\cite{mitchell2012partitioning}. 

The set of stable and unstable manifolds from the anchor points to the strong pips defines the boundary of a resonance zone\cite{mackay1987resonances, lomeli2009resonance}. We are concerned with quantifying transport between these resonance zones. The resonance zones give us where to establish our initial conditions for the Monte Carlo simulations and the escape barriers to compute when a trajectory has escaped. Once a trajectory has escaped from the outer period-1 resonance zone, it will never re-enter; some points that escape the inner period-10 resonance zone will eventually be remixed.

A lobe is a closed region of the phase space between two segments of the stable and unstable manifolds defined by two adjacent primary intersection points\cite{wiggins1990geometry, romkedar1994homoclinic}. As lobes are mapped forward, they are stretched and folded into and out of the resonance zone. Lobes that transport material out of the resonance zone upon forward iteration are referred to as ‘escape lobes’\cite{mitchell2003escape1, mitchell2003escape2}. Using the escape lobes, one can easily compute and estimate the escape rate $\gamma$ of material from a resonance zone\cite{mackay1987resonances}. The lobe area estimate computes the escape rate by taking the ratio of the escape lobe area and the area of the resonance zone as shown in \ref{escape_approximation}. This computes the ratio of material contained within the resonance zone with material that will be transported out of the resonance zone. This method works especially well when lobes are long and skinny as we observe in this system.

\begingroup
\setlength{\abovedisplayskip}{3pt}
\setlength{\belowdisplayskip}{3pt}
\begin{equation}
\label{escape_approximation}
    \gamma \approx \frac{A_{escape}}{A_{resonance}}
\end{equation}
\endgroup

\section{Magnetic Field Line Escape Rate}

A defining characteristic of moving away from integrability in Hamiltonian systems is that trajectory escape is now expected. As such, it is a benchmark for most techniques in the analysis of chaos \cite{cvitanovic2025chaosbook, wiggins1992chaotic}. Magnetic field line escape is particularly important in the physics of fusion reactors, as, to zeroth order, a field line can be thought of as an electron trajectory. Electrons carry the most thermal energy, and thus a large escape rate spells trouble for the longevity of the reactor. This was the concern presented in early studies of chaotic magnetic fields in fusion plasmas \cite{rechester1978electron}, where the chaos was assumed so strong that all traces of integrability was lost. Of course, as can be seen in Fig. \ref{fig:poincare}, this is not the case. In general \cite{evans2007separatrix, kim2015ideal}, there \emph{is} still remaining integrability even within a chaotic magnetic sea. To probe the effects of this additional structure, we first divide the trajectory space via the resonance zones of the period-1 orbit at the magnetic equilibrium x-point (see Fig.\ref {fig:equilpoincoord}) and the final period-10 island chain. 

\subsection{Monte Carlo Determination of Escape Rate}
In order to explore the consequences of the layered island chains, a Monte Carlo simulation of a large ensemble of trajectories is conducted. The simulation is undertaken by first seeding a uniform distribution of initial field line positions over the entire poloidal extent of the shot \#171491 equilibrium, and radially from $\psi^{eq}_n=0.8$ to $\psi^{eq}_n=1.0$. This allows complete coverage of all considered topological structures, and saves simulation time which would have been spent spreading the ensemble poloidally around the tokamak. For the results in this paper, $1,960,000$ points made up this initial uniform distribution, and each one was used to integrate eqn \eqref{field_line_eqns} for 100 total toroidal turns, so that each trajectory could (maximally) traverse the period-10 island chain structure 10 times. 

\begin{figure}[t]
\centering
\includegraphics[width=\columnwidth]{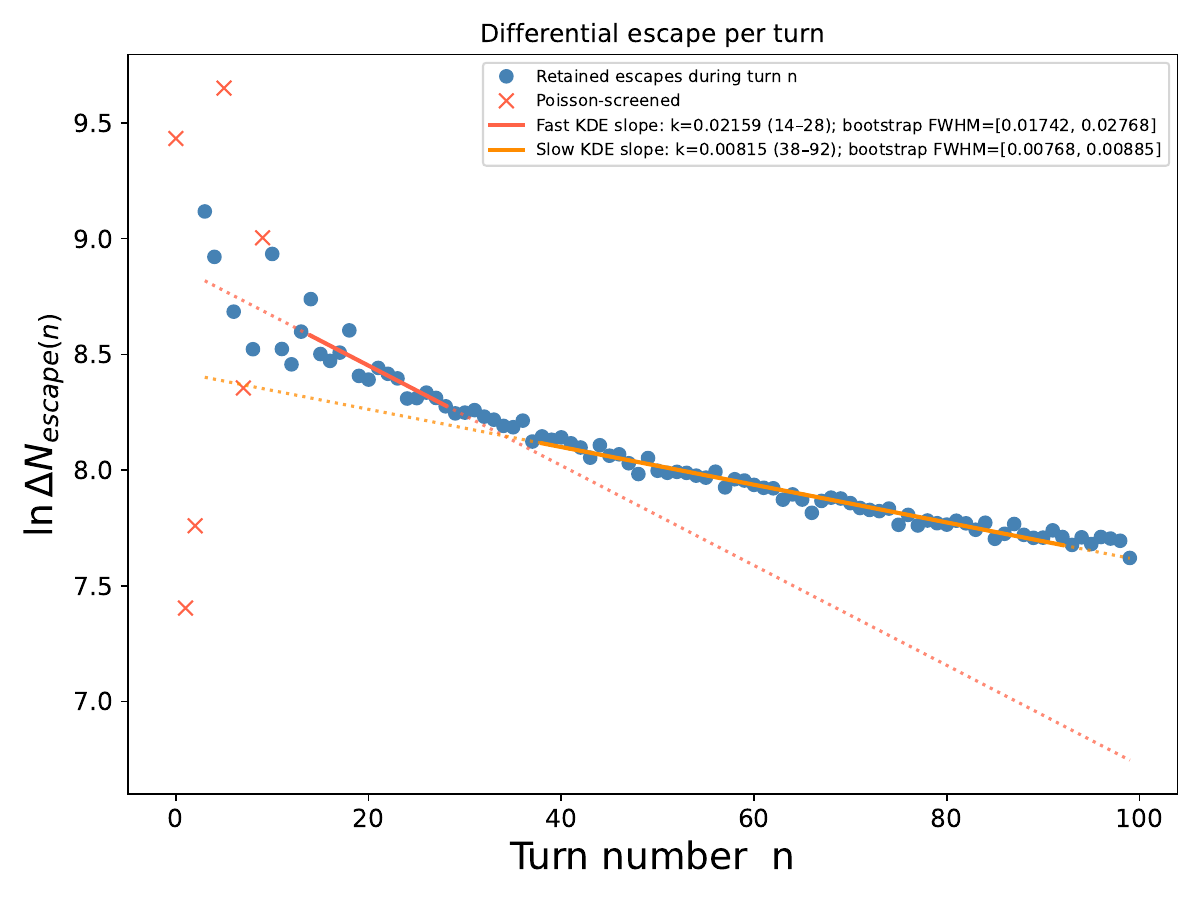}
\caption{\label{fig:monteone} Monte Carlo simulation of the escape rate, defined as crossing the single boundary created by the outer, period-1 tangle. The simulation was seeded with 1,960,000 points uniformly distributed in the region $\{0.8\leq\psi_N\leq 1 \hspace{3pt},0\leq \theta\leq 2\pi\}$ in order to fully cover the outermost period-10 island chain. The bi-exponential escape rate is attributed to the transport barrier established by the island chain's stable and unstable manifolds \cite{sattari2017periodic, wiggins1992chaotic}. Here, the Poisson-screened points are points excluded from the fitting process entirely because they failed the loose Poisson deviance test $d=2*[\mathrm{count}*\mathrm{ln}(\mathrm{count}/\lambda)-\mathrm{count}+\lambda]<35$. 35 is chosen as an arbitrary large number so that only those points with extreme departures are excluded from the fitting process.}
\end{figure}

\begin{figure}[t]
\includegraphics[width=\columnwidth]{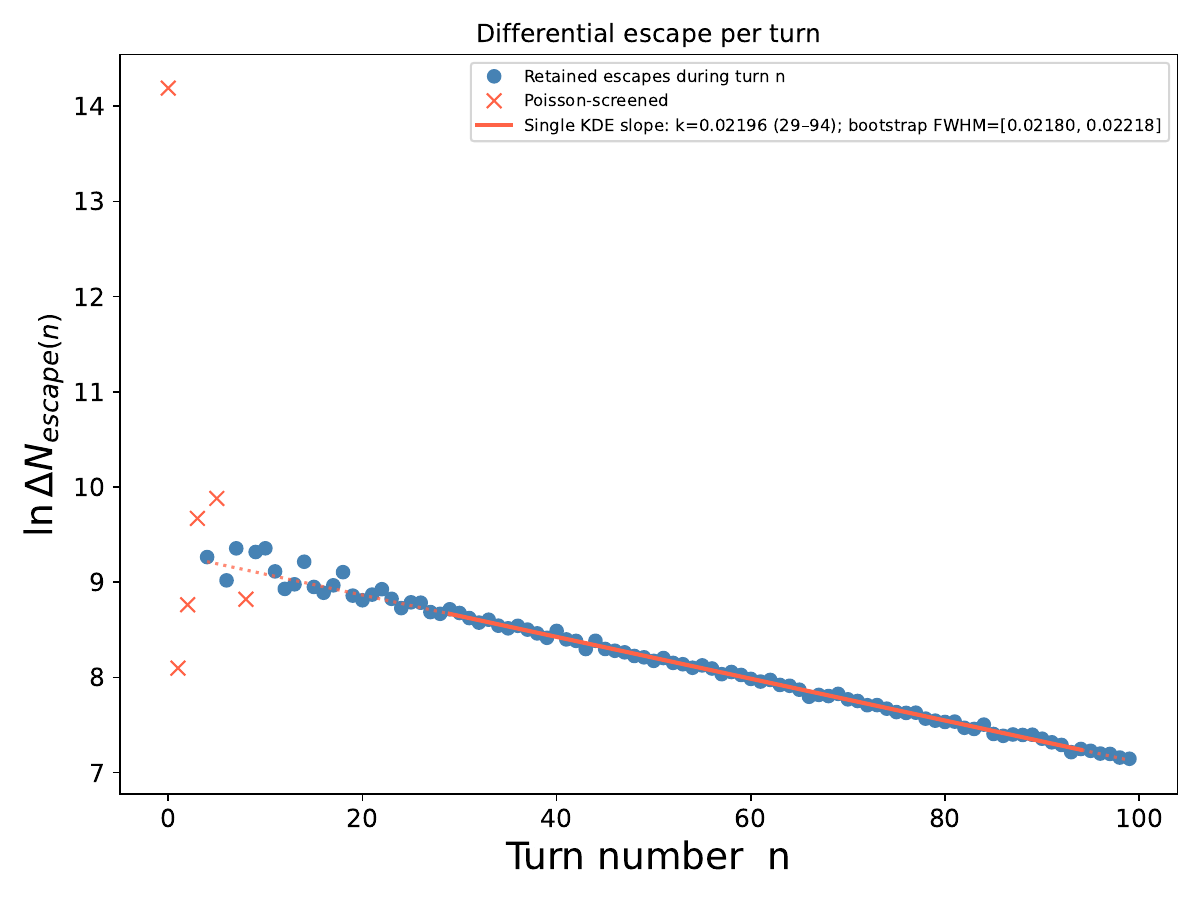}
\caption{\label{fig:montetwo} Monte Carlo simulation of escape rate defined as crossing either the boundary created by the outer, period-1 tangle, or the one created by the inner period-10 tangle. The simulation was seeded with 1,960,000 points uniformly distributed in the region $\{0.8\leq\psi_N\leq 1 \hspace{3pt},0\leq \theta\leq 2\pi\}$ in order to fully cover the outermost period-10 island chain. Here, the Poisson-screened points are points excluded from the fitting process entirely because they failed the loose Poisson deviance test $d=2*[\mathrm{count}*\mathrm{ln}(\mathrm{count}/\lambda)-\mathrm{count}+\lambda]<35$. 35 is chosen as an arbitrary large number so that only those points with extreme departures are excluded from the fitting process.}
\end{figure}

\begin{figure}[t]
\includegraphics[width=\columnwidth]{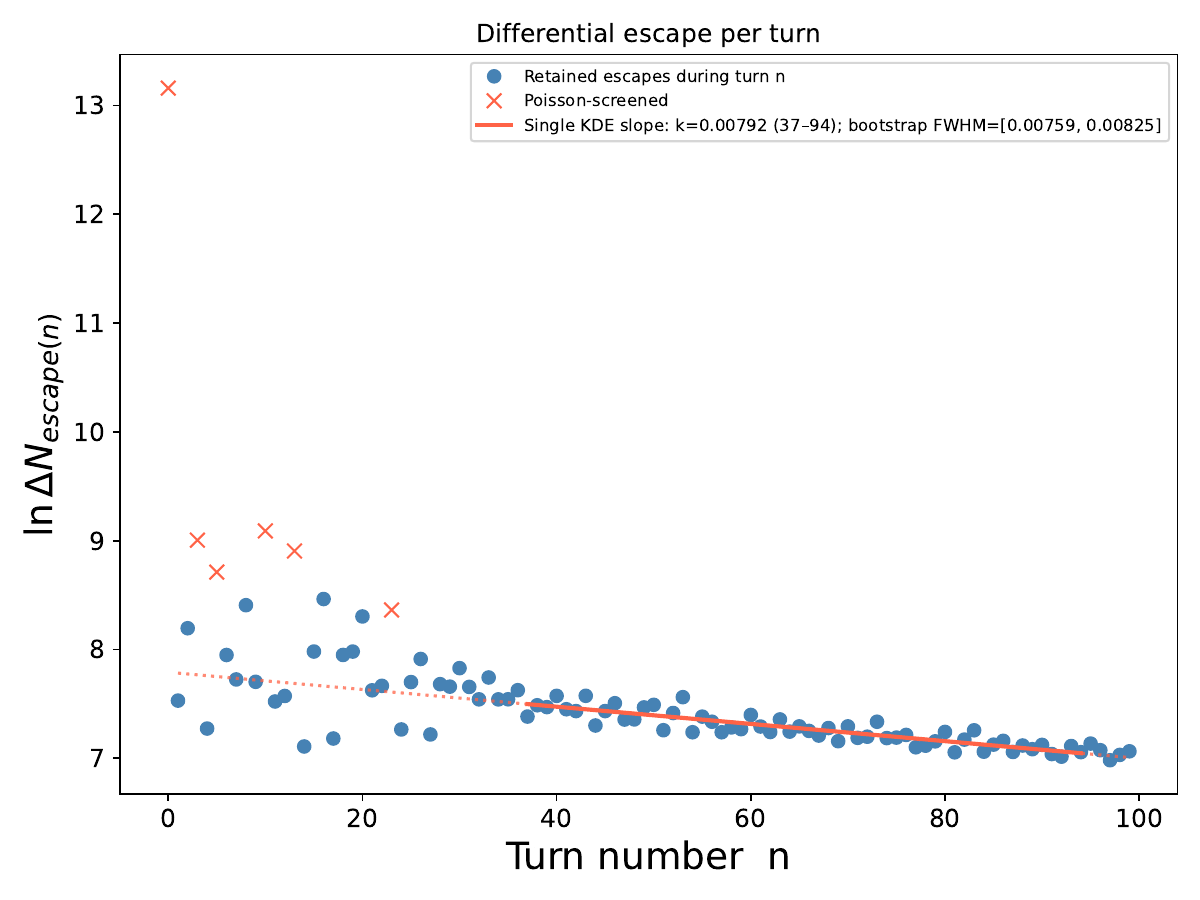}
\caption{\label{fig:monteinner} Monte Carlo simulation of escape rate defined as crossing outside the inner period-10 tangle. The simulation was seeded with 1,960,000 points uniformly distributed in the region $\{0.8\leq\psi_N\leq 1 \hspace{3pt},0\leq \theta\leq 2\pi\}$ in order to fully cover the outermost period-10 island chain. Here, the Poisson-screened points are points excluded from the fitting process entirely because they failed the loose Poisson deviance test $d=2*[\mathrm{count}*\mathrm{ln}(\mathrm{count}/\lambda)-\mathrm{count}+\lambda]<35$. 35 is chosen as an arbitrary large number so that only those points with extreme departures are excluded from the fitting process.}
\end{figure}

Tracking the position of the field lines as their trajectories evolve allows for an escape check. Essentially, a line must be drawn that defines when a trajectory has escaped the meaningful bounds of the system. In engineering-focused MAFOT operation, this line is defined as the reactor wall, but it is equally valid to set a more dynamical boundary. Here, we use the resonance zone boundaries---as defined in Sec. \ref{heteroclinic_tangles}---constructed via the algorithm defined in Appendix \ref{pip_algo}, and displayed in Fig. \ref{fig:stitchedmani}. These boundaries can be utilized in three different ways:
\begin{enumerate}
    \item Determining overall escape. This uses only the outer period-1 boundary (area enclosed by red and green manifold segments in Fig. \ref{fig:stitchedmani}), labeling all trajectories that pass over it as "escaped". The simulated escape pattern is visualized in Fig. \ref{fig:monteone}.
    \item Determining escape over the period-10 boundary. This uses just the period-10 resonance zone boundary (area enclosed by blue and gold manifold segments in Fig. \ref{fig:stitchedmani}), and labels any trajectories initiated within the boundary as "escaped" when they cross the boundary. The simulated escape plot is visualized in Fig. \ref{fig:monteinner}.
    \item Determining escape from the "hyperbolic" region. This utilizes both of the boundaries (the area between red and green along with blue and gold manifold segments in Fig. \ref{fig:stitchedmani}), labeling any trajectories that initiated between the period-10 island chain and the period-1 separatrix as "escaped" when they cross over either boundary. The simulated escape plot is visualized in Fig. \ref{fig:montetwo}.
\end{enumerate}
In each case, whether or not a trajectory has crossed a boundary is determined by a winding number point-in-polygon test \cite{alciatore1995winding, hormann2001point}.

The striking feature that is revealed through these simulations is that when only the outer boundary is used, the escape rate has a distinctly bi-exponential character (Fig. \ref{fig:monteone}.) This mirrors the results of earlier work which studied the nested island structure of the H\'enon map \cite{sattari2017periodic}. Here, as in Ref. \cite{sattari2017periodic}, the nested island chains are partitioning the dynamics, establishing a region of "fast" escape, and one of "slow" escape. This is an especially important realization when put into the context of classical analyses of chaotic fields in fusion reactors \cite{rechester1978electron, chirikov1979universal}, which usually considered the anti-integrable limit and predicted strong escape everywhere. As seen from the dual-boundary simulation (Fig. \ref{fig:montetwo}), this assumption is only defensible in the outer, colder edge of the reactor, making the prediction of strong field line escape leading to electron temperature profile flattening \cite{rechester1978electron} much less disastrous.  

Fundamentally, the dual-boundary simulation relies upon the assumption that the escape rate of those trajectories which cross into the period-10 resonance zone is so small that as far as the outer chaotic region is concerned, those trajectories are trapped. This is particularly useful as the period-10 resonance zone boundary is the last location where evidence of integrability appears in Poincar\'e sections (see Fig. \ref{fig:poincare}). By stripping all trajectories which cross into the period-10 resonance zone, the fully chaotic outer region's escape statistics are revealed to be a single exponential: a strong piece of numerical evidence that the chaotic sea past the period-10 boundary is hyperbolic. 

\subsection{Lobe Dynamics Calculation of Escape Rate}
\begin{figure}[ht]
    \centering
    \includegraphics[width=\columnwidth]{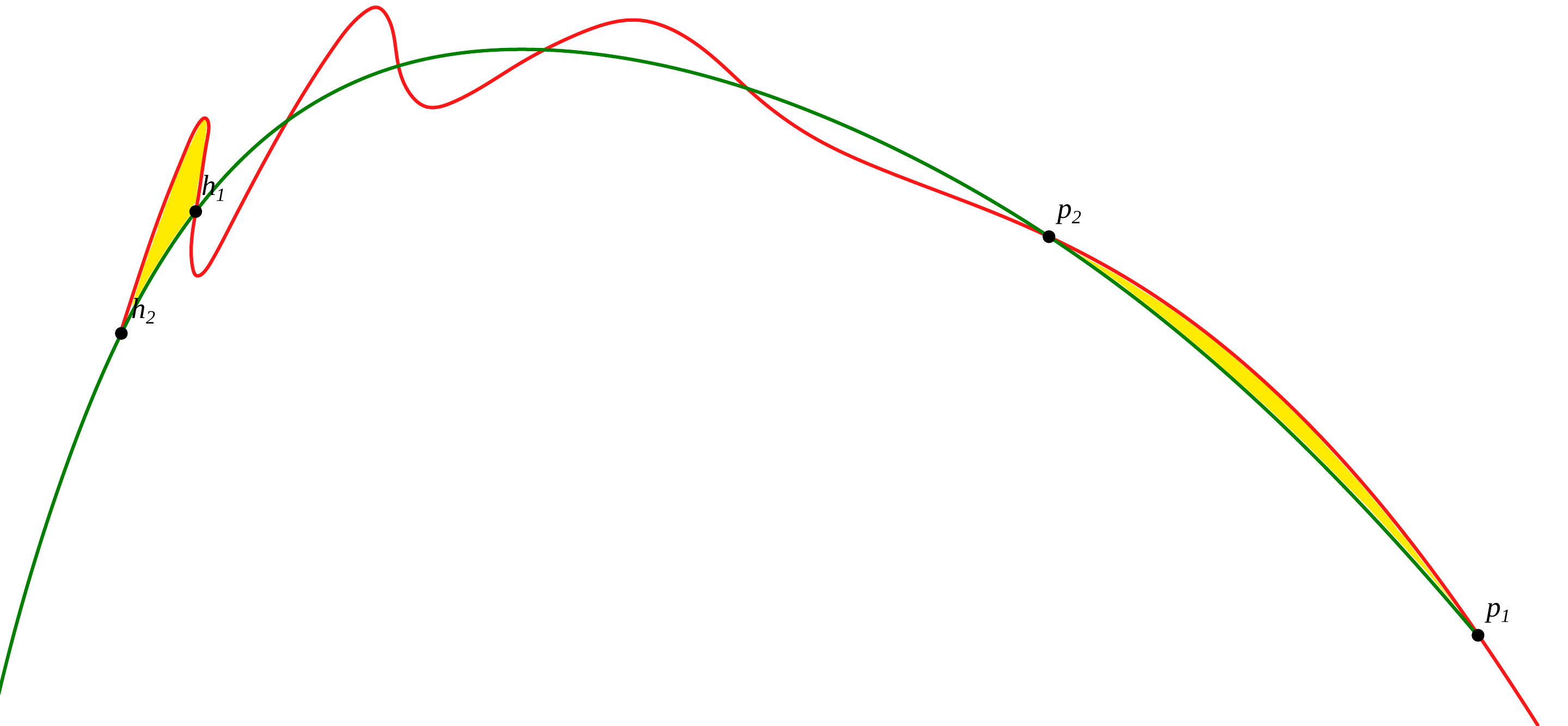}
    \caption{A visualization of transport of phase space through the outer period-one tangle. As an example, two lobes are filled in; the lobe furthest right is the initially considered ensemble, and the lobe furthest left is the result of mapping the points $p_1$ and $p_2$ fully around the tokamak. Note that the mapped lobe is the third in the chain; this is directly due to the $C_3$ symmetry of the perturbing magnetic field.}
    \label{fig:lobe_turnstile}
\end{figure}

Although Monte Carlo methods clearly demonstrate the various transport regimes present in a system strongly under the influence of integrability, it is still a long, largely random numerical integration. Fortunately, the basic mechanism may also be explored through the independent framework of lobe dynamics \cite{wiggins1992chaotic, meiss1992symplectic}. Turnstile flux has recently been applied explicitly as a quantitative measure of chaotic magnetic transport in magnetic-confinement configurations \cite{Smiet2025}. Within this framework, the transport across the resonance zone boundary which defines "escape," is governed entirely through the turnstile mechanism explained in Section \ref{heteroclinic_tangles}. The actual transport through the outer period-1 resonance zone in Shot \#171491 is visualized in Fig. \ref{fig:lobe_turnstile}. Note that instead of the traditional picture \cite{wiggins1992chaotic}, the lobe dynamics here transport the enclosed area \textit{three} lobes forward. This is due to the $C_3$ symmetry in the magnetic field perturbation. Effectively, when taking into account the whole fusion reactor (which is the correct strategy when dealing with heat load management), the n=3 external perturbation triples the escape rate one would expect when looking at a two-dimensional turnstile.

To calculate this predicted transport, eqn. \eqref{poincare-cartan} is calculated for a single lobe (denoted $L$).  This "area" element is then divided by the same quantity calculated for the "area" between the two resonance zone boundaries (denoted $R$) depicted in Fig. \ref{fig:stitchedmani} \cite{mackay1984transport, meiss1992symplectic}, the value is then multiplied by $3$ to account for the $C_3$ symmetry

\begin{equation}\label{lobe_escape}
    \gamma_{\mathrm{lobe}}
\approx
\frac{3\mu_\omega(L)}
{\mu_\omega(R)}.
\end{equation}

This procedure yields an escape rate, $\gamma_{\mathrm{lobe}}$, of $\sim 0.021$, a very close match with the full Monte Carlo, available in a fraction of the time, and tied directly to established dynamical systems theory. This is the primary result of the paper. 

The explanation we adopt for the close agreement between the single boundary lobe dynamics estimate, and the dual boundary system’s asymptotic escape rate is that the quasi-steady distribution of the dual boundary system ceases to be uniformly distributed. This violates the conditions needed for the simple lobe estimate ($\gamma\approx 3*(\mu_{\omega}(L_o)+\mu_{\omega}(L_i))/\mu_{\omega}(R)$ where $L_{o/i}$ is the single lobe measure of the outer, or inner PIP boundary) to apply exactly. Applying this estimate to the dual boundary system results in an escape rate which is 28\% too large. Initializing a Monte Carlo simulation with trajectories uniform in the invariant measure reproduces this enhanced escape as a transient before rapidly settling to the rate reported above. The implications of this nonuniform quasi-steady distribution on the system as a whole will be the subject of future investigations.

Note that despite the Monte Carlo simulation being initialized from a uniform $(\psi_N,\theta)$ distribution, which can affect mean escape times and transient effects, the asymptotic escape rate is a property of the map's underlying invariant structure, independent of the initial ensemble \cite{altmann2013leaking}. It is therefore valid to compare the small-flux approximation Eq. \eqref{lobe_escape} of the turnstile escape rate to the independent Monte Carlo escape rate. No identification between the Monte Carlo sampling measure $d\psi_Nd\theta$ and the invariant symplectic measure $(\mu_\omega)$ is assumed in this comparison.

\section{Conclusions and Future Work}

Magnetic field-line transport in perturbed tokamak plasmas is frequently interpreted through the lens of global stochasticity and diffusive transport. The present work demonstrates that such descriptions can overlook critical dynamical structures that continue to regulate transport even after resonant overlap has generated a substantial chaotic layer. Using magnetic field-line tracing in DIII-D discharge 171491, we identified a hierarchy of nested resonant island chains and their associated stable and unstable manifolds, revealing a transport topology far richer than that implied by stochasticity alone.

The principal result of this work is the identification of nested tangles as dynamically significant transport barriers. By constructing invariant boundaries from primary intersection points of the manifold structures, we isolated the transport effects of an outer period-10 island chain embedded within the stochastic edge region. Monte Carlo simulations showed that field-line escape through the perturbed separatrix exhibits a bi-exponential decay when only the outer transport boundary is considered. This behavior arises naturally from the existence of distinct transport domains: a rapidly escaping outer region and a sticky inner region whose transport is constrained by the island-chain manifold structure.

When the nested barrier is explicitly incorporated into the analysis, the escape process becomes approximately single exponential and is accurately described by turnstile transport through the outer homoclinic tangle. Furthermore, the measured escape rate agrees with a simple lobe-dynamics estimate derived from the symplectic geometry of the field-line map. This agreement establishes a direct connection between geometrical properties of the magnetic topology and macroscopic transport rates.

More broadly, these results support the view that transport in chaotic magnetic fields is organized by a finite set of invariant structures rather than by stochastic diffusion alone. The stable and unstable manifolds associated with resonant island chains form a transport skeleton that partitions phase space into regions with distinct dynamical behavior and characteristic escape times. Such structures provide a natural framework for understanding sticky transport, partial barriers, and non-diffusive field-line dynamics in magnetically confined plasmas.

The methodology developed here is general and extends beyond the specific DIII-D configuration examined in this work. Future investigations will incorporate plasma response effects, additional resonant island chains, and more realistic magnetic equilibria. The framework also opens the possibility of applying symbolic dynamics, topological entropy, and periodic-orbit theory to quantify transport in experimentally relevant fusion configurations. Ultimately, understanding and controlling these dynamical transport barriers may provide new pathways for optimizing resonant magnetic perturbations, tailoring edge transport, and improving the predictive capability of magnetic transport models in fusion plasmas.

\section*{Data Availability}

The data that support the findings of this study are available from the corresponding author upon reasonable request. The calculations were performed using a development version of MAFOT together with custom analysis scripts that are not currently available for public release. Additional information regarding the computational methods is available from the corresponding author upon reasonable request.

\section*{Acknowledgments}
This material is based upon work supported by the U.S. Department of Energy, Office of Science, Office of Fusion Energy Sciences, using the DIII-D National Fusion Facility, a DOE Office of Science user facility, under Award(s) DE-FC02-04ER54698, DE-SC0026433, and DE-FG02-05ER54809.

\appendix
\section{PIP Stitching Algorithm}\label{pip_algo}
Central to this work was the determination of appropriate boundaries which when crossed defines an escaped trajectory. As described in Section IV, these boundaries are formally the boundaries of the period-1 an period-10 resonance zones, which consist of alternating segments of the stable and unstable manifolds of the period-1 and period-10 hyperbolic periodic orbits, respectively.

Although at its heart, the method of constructing these segments consists simply of successive intersection checks, the chaotic nature of the system generating the manifolds introduces several nontrivial numerical difficulties. Namely:
\begin{itemize}
    \item An appropriate dynamical boundary must map each intersection point to the next intersection point defined by the order of the periodic orbit.
    \item Tracking intersections solely in one direction (forward or backward in "time") will always shrink one manifold segment, while allowing the other to stretch, and fold to extreme degrees. This not only makes finding the intersection more difficult, but the subsequent Monte Carlo simulation less robust. 
    \item As the choice of which initial intersection is used as the seed for a given boundary, many candidate intersections must be checked for numerical issues prior to settling on a final boundary.
\end{itemize}

 However, before dealing with any of the above issues, manifolds from which segments may be extracted must be generated. This is accomplished through the standard MAFOT framework \cite{wingen2009highresolution} which first extracts the stable, and unstable directions through a power method which applies the backward, and forward maps respectively to a small randomly-initialized vector which is renormalized after each iteration:
 \begin{equation}
     \mathbf{v}^{u/s}=\lim_{n\rightarrow\infty}|\mathbf{v}_0|\frac{(A^{\pm 1})^n(\mathbf{v}_0)}{||(A^{\pm 1})^n(\mathbf{v}_0)||}
 \end{equation}

 Over the course of the power method, a small initial segment of an invariant manifold, $\mathbf{x^{u/s}_{s}}$, is found which serves as the seed points for growing more of the manifold. This is accomplished by taking $\mathbf{v}^{u/s}$ and applying the appropriate map to it, renormalizing afterwards such that when adding $\mathbf{v}^{u/s}$ to a given fixed point, $\mathbf{x}_*=(A^{\pm 1})^{p}(\mathbf{x}_*)$, its image under the appropriate map has a magnitude within the range of $[5\times10^{-4},10^{-3}]$:

\begin{equation}
\begin{aligned}
& A^p(\mathbf{x}_*) = \mathbf{x}_*,
\
\mathbf{v}^\alpha = \epsilon_\alpha \hat{\mathbf{e}}_\alpha,
\
\mathbf{x}_{\mathrm{s}}^\alpha = \mathbf{x}_* + \mathbf{v}^\alpha, \
 \epsilon_\alpha \ \text{is chosen s.t.}\\
& 5\times 10^{-4}
\le
\left\|
A^{\pm 2p}\!\left(\mathbf{x}_{\mathrm{s}}^\alpha\right)
-
\mathbf{x}_{\mathrm{s}}^\alpha
\right\|_2
\le
10^{-3}, \ \alpha\in[u,s].
\end{aligned}
\end{equation}
 
 $A^{\pm 2p}(\mathbf{x}_{s})-\mathbf{x}_{s}$ is then taken as the \textit{Fundamental Segment} \cite{mitchell2009topology} which is subsequently used to grow either the unstable ($\mathcal{M}_u$), or stable ($\mathcal{M}_s$) manifold. This is accomplished by taking a small step along the Fundamental Segment and iterating this sub-segment with the appropriate map two times the period of the considered fixed point (the multiple of two guarantees branch consistency). After the fundamental segment is exhausted, should more of the manifold be needed, the process is restarted with the iteration count now being $2*k*\text{period}$, where $k$ starts at $1$ and is incremented each time the Fundamental Segment is exhausted. Without careful management, spacing between iterated points can become large, resulting in non-smooth manifolds, this leads directly to MAFOT's manifold growth constraints which force the developed manifold to maintain (relative) numerical smoothness by adapting the step size along the Fundamental Segment to prevent new segments from being too long, or too short. The entire algorithm can then be put in set-builder notation as eqn. \eqref{manigrowth_constraint}:

\begin{widetext}
\begin{equation} \label{manigrowth_constraint}
\begin{split}
    \mathcal{M}^{x_*}_{u/s}
    =
    \left\{
    A^{\pm 2kp}(\mathbf{x}^{\alpha}_s+t_i^{(k)}d\mathbf{a}^{\alpha})
    \;\middle|\;
    k=1,2\ldots,\ t_i^{(k)}\in T_k,\
    \mathbf{x}^{\alpha}_s=\mathbf{x}_*+\mathbf{v}^{\alpha},\ d\mathbf{a}^{\alpha}=A^{\pm 2p}(\mathbf{x}^{\alpha}_s)-\mathbf{x}^{\alpha}_s,\
    5{\times}10^{-4}\le\|d\mathbf{a}^{\alpha}\|\le10^{-3}
    \right\},
    \\
    T_k=\left\{0=t_0<\cdots<t_{N_k}=1
    \;\middle|\;
    \left\|A^{\pm 2kp}(\mathbf{x}^{\alpha}_s+t_i d\mathbf{a}^{\alpha})-A^{\pm 2kp}(\mathbf{x}^{\alpha}_s+t_{i-1}d\mathbf{a}^{\alpha})\right\|\le5{\times}10^{-3}
    \right\}, \ \alpha\in[u,s].
\end{split}
\end{equation}
\end{widetext}

\eqref{manigrowth_constraint} is carried out for each manifold emanating from each point in each considered periodic orbit until there is significant structure formed (see fig. \ref{fig:fullmani}). From there, the outermost set of manifolds is chosen for a given orbit, and the intersecting stable/unstable pair with the least initial curvature is chosen. Utilizing the necessary and sufficient conditions proven in \cite{naik2017computational, gavrilova2000reliable}:
\begin{theorem}
If the segments $[(x_1,y_1),(x_2,y_2)]$ and
$[(x'_1,y'_1),(x'_2,y'_2)]$ intersect, then we must have
\begin{equation}
\begin{aligned}
&\left((y_2-y_1)x'_1-(x_2-x_1)y'_1-x_1y_2+y_1x_2\right) \\
&\qquad \times
\left((y_2-y_1)x'_2-(x_2-x_1)y'_2-x_1y_2+y_1x_2\right)
\le 0 .
\end{aligned}
\end{equation}
\end{theorem}

\begin{theorem}
The segments $[(x_1,y_1),(x_2,y_2)]$ and
$[(x'_1,y'_1),(x'_2,y'_2)]$ intersect if and only if
\begin{equation}
\begin{aligned}
&\left((y_2-y_1)x'_1-(x_2-x_1)y'_1-x_1y_2+y_1x_2\right) \\
&\qquad \times
\left((y_2-y_1)x'_2-(x_2-x_1)y'_2-x_1y_2+y_1x_2\right)
\le 0 ,
\end{aligned}
\end{equation}
and
\begin{equation}
\begin{aligned}
&\left((y'_2-y'_1)x_1-(x'_2-x'_1)y_1-x'_1y'_2+y'_1x'_2\right) \\
&\qquad \times
\left((y'_2-y'_1)x_2-(x'_2-x'_1)y_2-x'_1y'_2+y'_1x'_2\right)
\le 0 .
\end{aligned}
\end{equation}
\end{theorem}

Theorem 1 is checked first for each segment pair in the chosen manifold pair, it gives a necessary condition for an intersection between two segments. If Theorem 1 is passed, Theorem 2 (the sufficient condition) is checked, and then if that too passes the intersection point is calculated. Importantly, as these calculated intersection points accrue both machine, and manifold-discretization error, they can only be trusted up to about one map (forward, or backward) iteration before drifting noticeably off of the manifold. To remedy this, along with the intersection point, the two points on the "other" side of the intersection point are tracked. These points are in $\mathcal{M}^{x_1}_{u/s}$ and so they remain on the manifolds of subsequent points in the periodic orbit under map iteration. This directly addresses the first difficulty mentioned in this appendix by building into the algorithm intersections related by iteration. By finding a first intersection and its flank points, then iterating the considered manifold pair from the considered periodic point pair to the flank points the next intersection will be found and will be related by iteration to the previous intersection up to round-off, and manifold discretization error. To address the other difficulties, which portend to the increasing complexity of manifold topology as the iteration-related PIP boundary is grown, two actions are taken. First, the initial distance along each manifold from the intersection point to that manifold's respective flank point is tracked. After each iteration, and intersection calculation, the "stubs" after the intersection are refilled to the tracked flank distance using the appropriate manifold in $\mathcal{M}_{u/s}^{\mathbf{x}}$, this prevents a shrinking manifold from becoming too short through accumulation of numerical error to find an intersection. Second, after assembling the first PIP boundary from a specific choice of initial intersection, the volume and total arclength is checked, and then another boundary is assembled choosing either a different combination of forward and backward iterations, or another initial intersection point. This is repeated until all combinations are exhausted, and then the combination that yields the minimal volume and arclength is chosen. By checking all these combinations, the least complicated boundary can be determined, and as the initial manifold pair is chosen to minimize initial curvature, the number of configurations to check is often rather small.

\section{Bi-Exponential Fitting Algorithm}

The analysis estimates the exponential decay rate of the escape-turn distribution conditional on escape within the analyzed turn window. It is not the unconditional probability of escape, the fraction remaining at the final turn, or a population hazard estimated from all launched trajectories. Therefore, nonescaping trajectories need not be treated as an explicitly separate population within this analysis. With that in mind, in order to validate the statements implied by fig. \ref{fig:monteone} and fig. \ref{fig:montetwo} it is necessary to establish both
\begin{itemize}
    \item The bi-exponential character in fig. \ref{fig:monteone} is not a fitting artifact
    \item Adding the period-10 loss-channel truly extracts single-exponential escape dynamics in fig. \ref{fig:montetwo}
\end{itemize}

Checking these points is best approached through a statistical framework. Effectively, the problem is formulated as the result of a draw from a multinomial distribution with a fixed number of escapes $N$ over the course of $n$ toroidal turns:
\begin{equation}
\begin{split}
    (e_1,e_2,\ldots,e_n)\sim\text{Multinomial}\{N;\pi_1,\pi_2,\ldots,\pi_n\}\\
    \pi_i=\operatorname{Pr}(T=i \hspace{5pt} | \hspace{5pt} T\leq n),\quad \sum_{i=1}^n\pi_i=1
\end{split}
\end{equation}
Where $\pi_i$ is the is the probability of escaping at the $i^{th}$ turn around the tokamak. As the data being fit is count data, a multinomial regression is the obvious tool. Unfortunately, with the sheer number escapes present in the dataset, even small systematic departures from an ideal exponential dominate multinomial likelihood, so a more robust working likelihood must be selected. In our case the Negative Binomial distribution is the chosen regression tool which will be used in all places except fit boundary selection. The specific regression model is 

\begin{equation}\label{NB_model}
Y_n \sim \operatorname{NB}(\mu_n,\alpha),
\hspace{3pt}
\operatorname{E}(Y_n\mid\mu_n)=\mu_n,
\hspace{3pt}
\operatorname{Var}(Y_n\mid\mu_n)
=
\mu_n+\alpha\mu_n^2.
\end{equation}
The later utilized model selection metric will be calibrated to account for the working nature of the Negative Binomial regression on this dataset.

Boundary selection follows the algorithm developed by Deshmukh et. al. where a weighted histogram of Poisson Regression along with Kernal Density Estimation \cite{deshmukh2021scaling} produces fitting intervals for each of the $x$ most "prominent" modes of the weighted slope distribution function. Here, prominence is measured via the peak prominence$=\text{peak height}-\text{adjacent saddle level}$. This naturally deweights local peaks that are actually shoulders in the larger distribution. 

In simple datasets, Deshmukh's method is sufficient to extract relevant two-feature characteristics of point clouds. The escape data generated by MAFOT does not yield to just Deshmukh's method. As a large percentage of the initial trajectories fall behind the period-10 resonance zone boundary, the "fast" escape rate is dominant for a relatively small number of toroidal turns. This is heavily penalized in Deshmukh's method as the histogram weighting is calculated as
\begin{equation}\label{kde_weight}
    w\propto\frac{(\text{fit length})^p}{(\text{fit error})^q}
\end{equation}
Where we have chosen $q=p=1$. eqn \eqref{kde_weight} grants preference to accurate fits which were done over a large interval. The "large interval" factor is no available in the short fast region.

To address the down-weighting of the fast region, it is necessary to determine whether it is true down-weighting of a short--but relevant--region, or an aberration generated by the Monte Carlo simulation. To test for this, the Bayesian Information Criterion ($\mathrm{BIC}$) is calculated for for both a single, and a double exponential model through Negative Binomial Regression. From Schwarz \cite{schwarz1978estimating} the model which should be selected for the data has the smallest $\mathrm{BIC}$ (the Schwarz criterion is half the negation of $\mathrm{BIC}$, however, a minimization criteria has long been standard \cite{katz1981criteria}. This, along with a heuristic measure from Kass and Raftery \cite{kass1995bayes}, yields the following check on whether a de-weighted mode from Deshmukh's method is real signal or noise:
\begin{equation}\label{BIC_condition}
    \Delta\mathrm{BIC}
    \equiv \mathrm{BIC}_1-\mathrm{BIC}_2=2(l_2-l_1)-2\log(m)>10
\end{equation}
Where $l_i$ is the maximized negative binomial log likelihood for either the single, or double exponential, and $m$ is the number of retained turn bins. eqn. \eqref{BIC_condition} is derived from the following one- and two-exponential models:
\begin{equation}\label{exp_models}
    \begin{split}
        \mu_n=A\exp(-\kappa n)\\
        \mu_n=A_f\exp(-\kappa_fn)+A_s\exp(-\kappa_sn)
    \end{split}
\end{equation}
The single exponential model fits 3 parameters: $(A,\kappa,\alpha)$, whereas the double exponential model fits 5 parameters: $(A_f,A_s,\kappa_f,\kappa_s,\alpha)$. \eqref{BIC_condition} is immediate from the definition of the Bayesian Information Criterion: $\mathrm{BIC}=k\log(m)-2l$, where $k$ is the number of fitted parameters.

Because the Negative Binomial model is used only as a working likelihood for the model decision, null datasets are generated from the fixed-total multinomial sampling model rather than from a negative-binomial model. To check Kass and Raftery's heuristic for our working Negative Binomial likelihood, a null calibration is conducted. The null hypothesis is that the conditional escape-turn distribution follows a single exponential profile. To test whether the decision rule would falsely select a double exponential, a second Multinomial distribution is generated by fitting the data from fig. \ref{fig:montetwo} with a single exponential model, normalized over the analyzed turns, and generating the probabilities of escaping at each turn $n$. This Multinomial distribution is then sampled with the same number of escapes as the original dataset. We do this 500 times, creating 500 test cases on which the $BIC$ for both a single, and a double exponential model are calculated. None of the 500 test cases generate a false positive under eqn. \eqref{BIC_condition}, and the one sided $95\%$ Clopper-Pearson upper confidence bound \cite{clopper1934use} is $0.6\%$. With this evidence we are confident that \eqref{BIC_condition} is a defensible condition to check if Deshmukh's method predicts meaningful departure from a single exponential model. 

With the statistical check calibrated, the analysis pipeline becomes 

\begin{enumerate}
    \item Check $\Delta \mathrm{BIC}$ for a given set of decay data, retain one mode from Deshmukh's method if less than 10, two if more
    \item Determine the fitting intervals via Deshmukh's method, and perform Negative Binomial Regression in that interval to obtain all statistically relevant escape rates
\end{enumerate}

The multinomial sampling model, Poisson interval regressions, and negative-binomial regression serve separate purposes in the analysis. They are not asserted to form a single common generative likelihood. The fixed-total multinomial model provides high-level statistical structure by which the $N$ total observed escapes get sorted into turn bins. Deshmukh's method, along with Poisson log-linear regression provide only prominence-sorted fitting intervals for slopes predicted via KDE analysis. Errors obtained here are not used as slope uncertainties, and Poisson likelihood is not used delineate between one- and two-exponential models. Negative Binomial working likelihood is used for finding the final regional rates, and testing global one- and two-exponential models. The additional dispersion parameter in Negative Binomial regression prevents high-count bins from receiving the excessive influence implied by $\operatorname{Var}(Y_n)=\mu_n$. This is the Negative Binomial parameterization described by Cameron and Trivedi \cite{cameron1986econometric}.

\begin{figure}[htbp]
    \centering
\begin{tikzpicture}[node distance=2cm]

\node (start) [startstop] {Observed Escape Histogram};
\node (in1) [process, below of=start] {Resample Turn Distribution With Replacement};
\node (pro1) [process, below of=in1] {Recompute $\Delta BIC$ for Single and Double Exponential Models};
\node (dec1) [decision, below of=pro1, yshift=-0.5cm] {Check $\Delta BIC$};

\node (pro2b) [process, below of=dec1,  xshift=-1.6cm ,yshift=-0.25cm] {Recompute multiple $x_r$ and $x_l$ from Deshmukh's Method};

\node (pro3) [process, right of=pro2b, xshift=2.0cm] {Redo Negative Binomial Regression in Each Interval};

\draw [arrow] (start) -- (in1);
\draw [arrow] (in1) -- (pro1);
\draw [arrow] (pro1) -- (dec1); 
\draw [arrow] (pro2b) -- (pro3);

\draw [arrow] (dec1.west)
    -- ++(-2cm,0)
    |- node[pos=0.25, anchor=south, rotate=90] {$\Delta BIC<10$: record slope, and fit interval then repeat}
    (start.west);

\draw [arrow] (dec1.south) -- node[anchor=south, xshift=-0.7cm, yshift=-0.1cm] {$\Delta BIC > 10$} (pro2b);

\draw [arrow] (pro3) |- node[pos=0.25, anchor=north, rotate=90] {log slopes and repeat} (start);
\end{tikzpicture}
\caption{Bootstrap pipeline}
    \label{fig:bootstrap}
\end{figure}
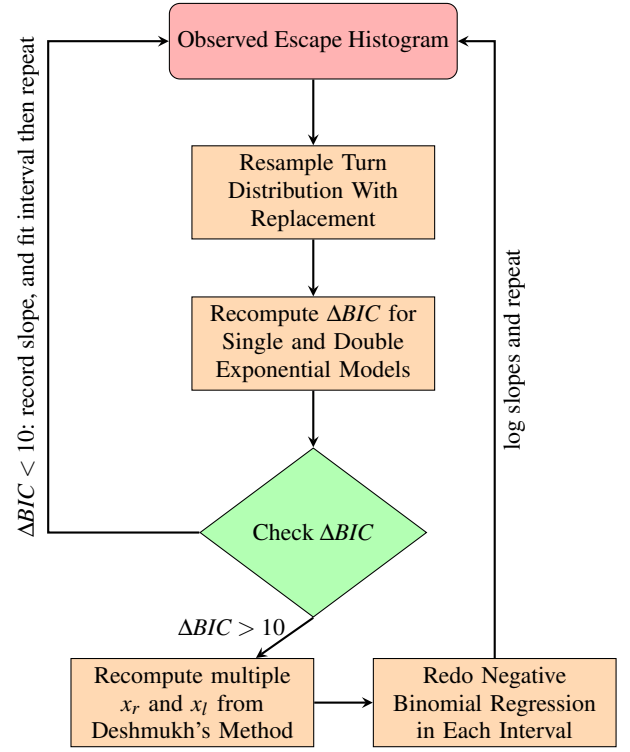

\begin{figure}[htbp]
    \centering
\includegraphics[width=\columnwidth]{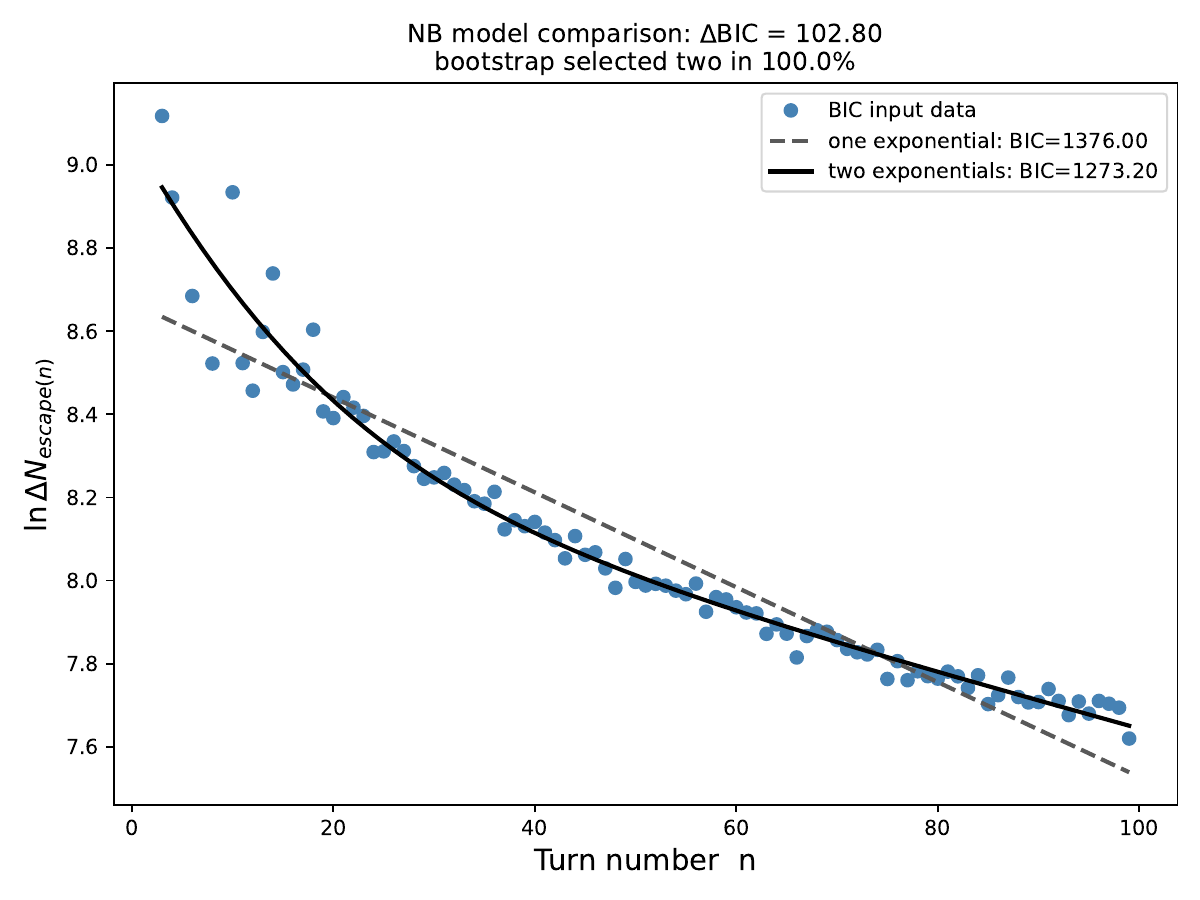}
\caption{Visualization of the Negative Binomial regression utilizing single- or double-exponential models. The $\Delta\mathrm{BIC}$ criteria (eqn. \eqref{BIC_condition}) is strongly in favor of a two exponential model. This conclusion is reproduced in 100\% of the bootstrapped escape histograms.}
    \label{fig:single_BIC}
\end{figure}

\begin{figure*}[htbp]
    \centering
\includegraphics[width=2\columnwidth]{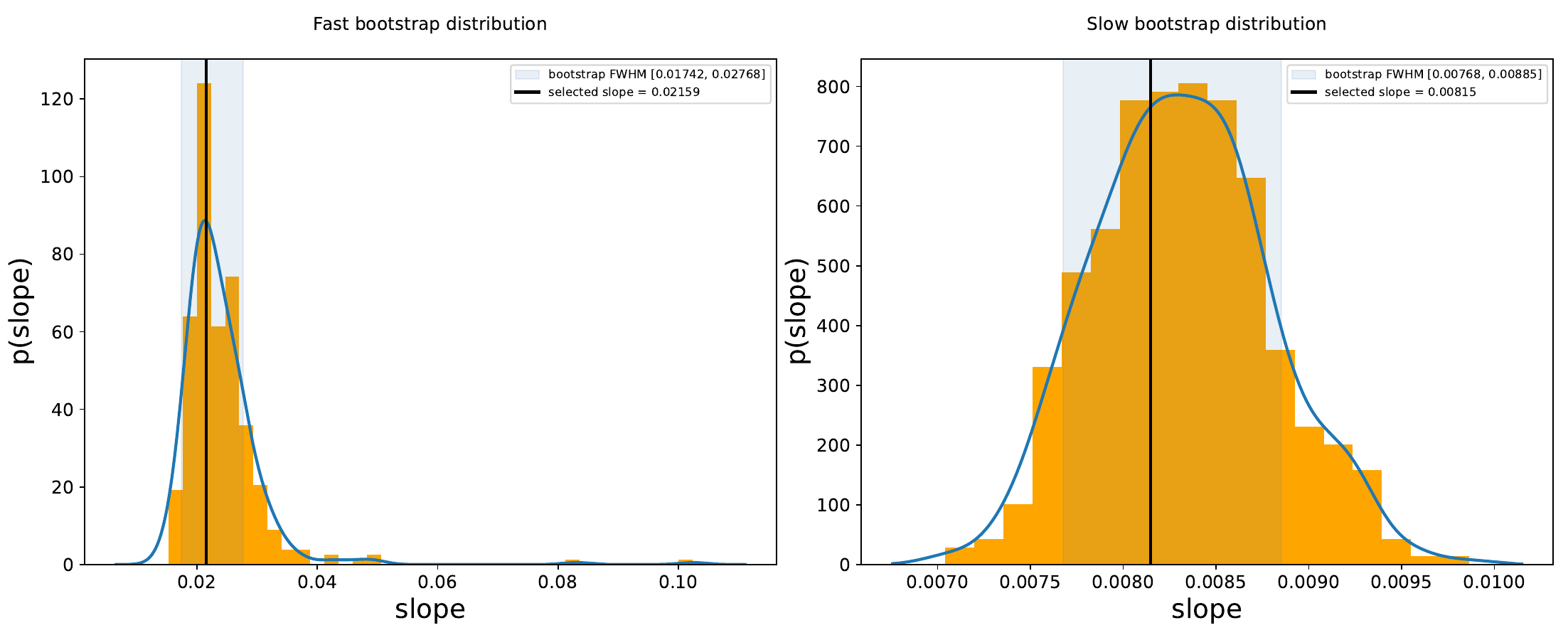}
\caption{Bootstrap distributions for the slope of both the fast and slow regions. These are generated following the procedure laid out in fig. \ref{fig:bootstrap} where new escape histograms are created by resampling with replacement, allowing for many additional runs of the statistical pipeline, including resampling of fitting intervals. Following Deshmukh's method, the full width half maximum of these distributions provide an approximate uncertainty range to the fitted slopes.}
    \label{fig:single_boot}
\end{figure*}

\begin{figure}[htbp]
    \centering
\includegraphics[width=\columnwidth]{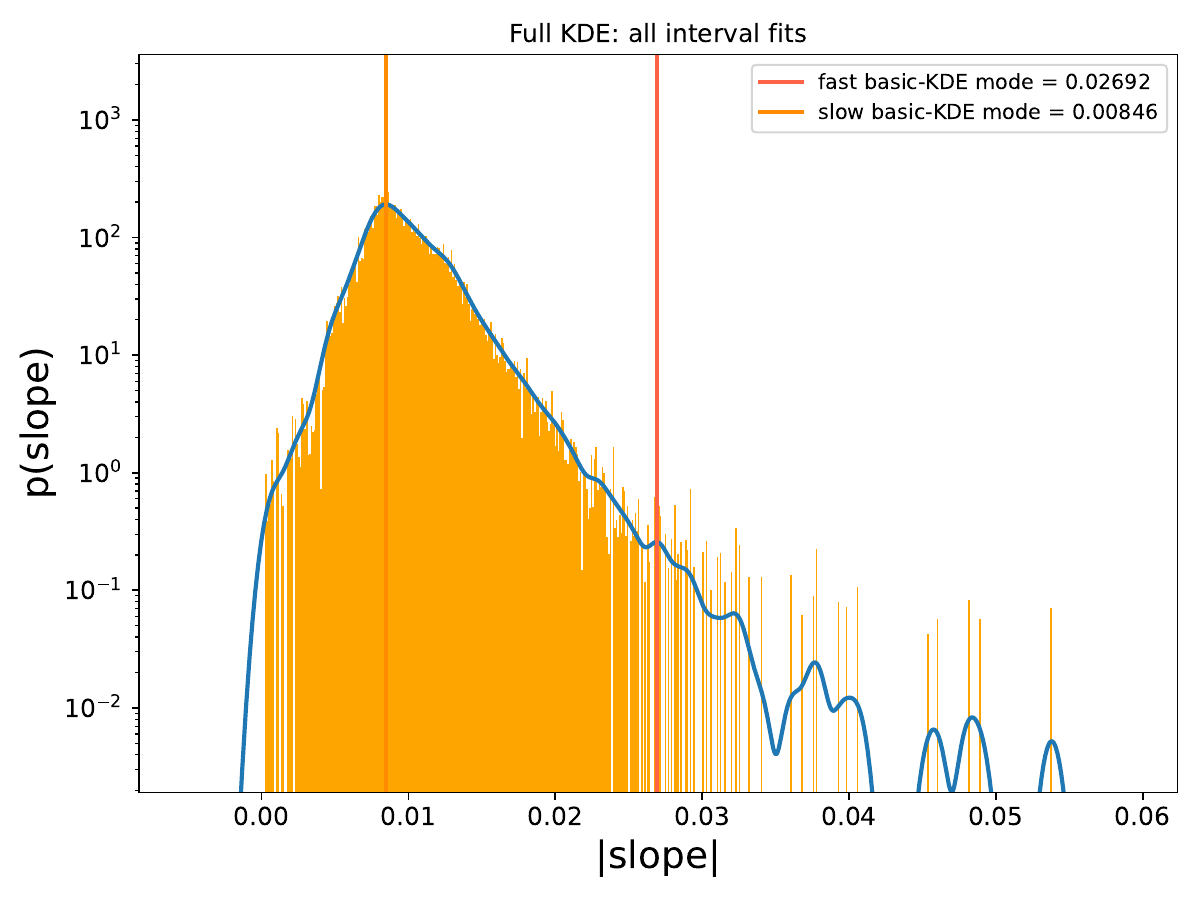}
\caption{Slope distribution of the nominal double-exponential escape profile. This was created via Deshmukh's method. Unfortunately, due to biasing against short, somewhat scattered regions, it is only really useful as a minor part of the fitting interval selection pipeline.}
    \label{fig:single_kde}
\end{figure}

\begin{figure}[htbp]
    \centering
\includegraphics[width=\columnwidth]{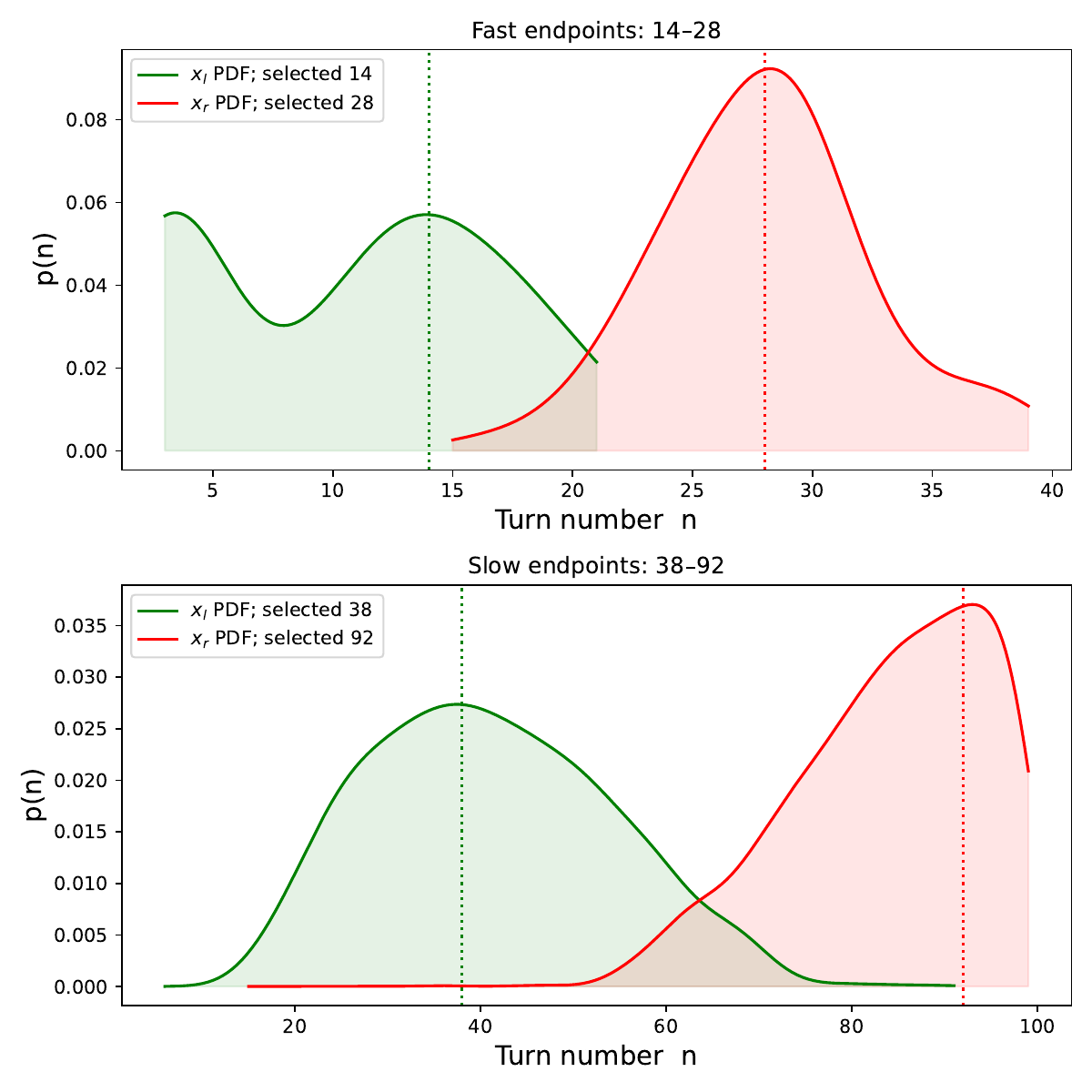}
\caption{Fitting endpoint distributions for the fast and slow decay regions generated by KDE via Deshmukh's method.}
    \label{fig:single_xlxr}
\end{figure}

Unfortunately, due to the aforementioned issue with placing the fast escape rate on the shoulder of the larger escape distribution, the uncertainties generated by Deshmukh's method are not trustworthy. We remedy this through Efron's Bootstrap method \cite{efron1979bootstrap}. During this procedure, the $N$ escapes are resampled with replacement and the intervals along with the slopes are recomputed. The uncertainties seen in fig. \ref{fig:monteone} and fig. \ref{fig:montetwo} are taken as the bootstrap distributions' full width half maximum to match the uncertainty quantity used in Deshmukh's method. The new distribution is generated through the process detailed in the flowchart in fig. \ref{fig:bootstrap}. The full results of this pipeline are in \cref{fig:single_BIC,fig:single_boot,fig:single_kde,fig:single_xlxr}.

Finally, to verify that adding the period-10 loss channel truly extracts a hyperbolic region with a single exponential decay, a power calibration on the sensitivity of the $\Delta\mathrm{BIC}$ gate is conducted. This is done by planting a slow component in the calibration's probability mass function (eqn. \eqref{power_PMF}) and checking if the $\Delta\mathrm{BIC}$ gate catches it. 

\begin{equation}\label{power_PMF}
    \begin{split}
        F_n=A\exp{(-\kappa_fn)},\quad S_{n;c}=F(c)\exp{(-\kappa_s(n-c))}\\
        p_n(c)=\frac{F_n+S_{n;c}}{\sum_{i=1}^n(F_i+S_{i;c})}\\
        c\in\{30,200,225,250,275,300\},\quad \kappa_s=0.0083
    \end{split}
\end{equation}

Here, $c$ is the crossover point which is varied over the course of the calibration, and $\kappa_s$ is the slow decay chosen to reflect what this system has already shown as a realistic value. The values of $c$ were chosen to start at a strong crossover location which mimics the conditions imposed by failing to fully remove integrable structure, and then extended to probe progressively weaker contributions. eqn. \eqref{power_PMF} generates another Multinomial distribution:
\begin{equation}\label{power_Multi}
    (e_1^*,\ldots,e_n^*)\sim \operatorname{Multinomial}(N;p_1(c),\ldots,p_n(c))
\end{equation}
Using eqn. \eqref{power_Multi}, the $\Delta\mathrm{BIC}$ procedure from above is repeated, with a detection of the planted second exponential being $\Delta\mathrm{BIC}\geq10$. Power (eqn. \eqref{power_def}) is calculated 50 times for each chosen $c$:
\begin{equation}\label{power_def}
    \mathrm{power}=\frac{\mathrm{number \hspace{3pt} detected \hspace{3pt} \Delta BIC\geq10}}{\mathrm{successful \hspace{3pt} fits}}
\end{equation}

The results are stored in table \ref{tab:power_calibration}, showing that the chosen model selection $\Delta\mathrm{BIC}$ criterion is powerful enough to detect departures from a single-exponential model which contribute at least 14.2\% of the expected count at the final analyzed turn. This provides strong evidence that the escape profile in fig. \ref{fig:montetwo} is statistically a single exponential profile at the resolution of this analysis.
\begin{table}[ht]
\centering
\caption{Power calibration for detecting the bi-exponential model as a function of $c$.}
\label{tab:power_calibration}
\[
\begin{array}{c|c|c|c}
c &
\text{slow fraction at final turn} &
\text{detected} &
\text{power} \\
\hline
30  & 72.8\% & 50/50 & 100\% \\
200 & 19.1\% & 50/50 & 100\% \\
225 & 14.2\% & 50/50 & 100\% \\
250 & 10.4\% & 36/50 & 72\% \\
275 &  7.5\% & 10/50 & 20\% \\
300 &  5.4\% &  0/50 & 0\%
\end{array}
\]
\end{table}

\newpage
\section*{Disclaimer}
This report was prepared as an account of work sponsored by an agency of the United States Government. Neither the United States Government nor any agency thereof, nor any of their employees, makes any warranty, express or implied, or assumes any legal liability or responsibility for the accuracy, completeness, or usefulness of any information, apparatus, product, or process disclosed, or represents that its use would not infringe privately owned rights. Reference herein to any specific commercial product, process, or service by trade name, trademark, manufacturer, or otherwise does not necessarily constitute or imply its endorsement, recommendation, or favoring by the United States Government or any agency thereof. The views and opinions of authors expressed herein do not necessarily state or reflect those of the United States Government or any agency thereof.

\begin{thebibliography}{57}%
\makeatletter
\providecommand \@ifxundefined [1]{%
 \@ifx{#1\undefined}
}%
\providecommand \@ifnum [1]{%
 \ifnum #1\expandafter \@firstoftwo
 \else \expandafter \@secondoftwo
 \fi
}%
\providecommand \@ifx [1]{%
 \ifx #1\expandafter \@firstoftwo
 \else \expandafter \@secondoftwo
 \fi
}%
\providecommand \natexlab [1]{#1}%
\providecommand \enquote  [1]{``#1''}%
\providecommand \bibnamefont  [1]{#1}%
\providecommand \bibfnamefont [1]{#1}%
\providecommand \citenamefont [1]{#1}%
\providecommand \href@noop [0]{\@secondoftwo}%
\providecommand \href [0]{\begingroup \@sanitize@url \@href}%
\providecommand \@href[1]{\@@startlink{#1}\@@href}%
\providecommand \@@href[1]{\endgroup#1\@@endlink}%
\providecommand \@sanitize@url [0]{\catcode `\\12\catcode `\$12\catcode `\&12\catcode `\#12\catcode `\^12\catcode `\_12\catcode `\%12\relax}%
\providecommand \@@startlink[1]{}%
\providecommand \@@endlink[0]{}%
\providecommand \url  [0]{\begingroup\@sanitize@url \@url }%
\providecommand \@url [1]{\endgroup\@href {#1}{\urlprefix }}%
\providecommand \urlprefix  [0]{URL }%
\providecommand \Eprint [0]{\href }%
\providecommand \doibase [0]{http://dx.doi.org/}%
\providecommand \selectlanguage [0]{\@gobble}%
\providecommand \bibinfo  [0]{\@secondoftwo}%
\providecommand \bibfield  [0]{\@secondoftwo}%
\providecommand \translation [1]{[#1]}%
\providecommand \BibitemOpen [0]{}%
\providecommand \bibitemStop [0]{}%
\providecommand \bibitemNoStop [0]{.\EOS\space}%
\providecommand \EOS [0]{\spacefactor3000\relax}%
\providecommand \BibitemShut  [1]{\csname bibitem#1\endcsname}%
\let\auto@bib@innerbib\@empty
\bibitem [{\citenamefont {Boozer}(2005)}]{boozer2005physics}%
  \BibitemOpen
  \bibfield  {author} {\bibinfo {author} {\bibfnamefont {A.~H.}\ \bibnamefont {Boozer}},\ }\bibfield  {title} {\enquote {\bibinfo {title} {Physics of magnetically confined plasmas},}\ }\href {\doibase 10.1103/RevModPhys.76.1071} {\bibfield  {journal} {\bibinfo  {journal} {Reviews of Modern Physics}\ }\textbf {\bibinfo {volume} {76}},\ \bibinfo {pages} {1071--1141} (\bibinfo {year} {2005})}\BibitemShut {NoStop}%
\bibitem [{\citenamefont {Meiss}(1992)}]{meiss1992symplectic}%
  \BibitemOpen
  \bibfield  {author} {\bibinfo {author} {\bibfnamefont {J.~D.}\ \bibnamefont {Meiss}},\ }\bibfield  {title} {\enquote {\bibinfo {title} {Symplectic maps, variational principles, and transport},}\ }\href {\doibase 10.1103/RevModPhys.64.795} {\bibfield  {journal} {\bibinfo  {journal} {Reviews of Modern Physics}\ }\textbf {\bibinfo {volume} {64}},\ \bibinfo {pages} {795--848} (\bibinfo {year} {1992})}\BibitemShut {NoStop}%
\bibitem [{\citenamefont {Kuramoto}(1978)}]{kuramoto1978diffusion}%
  \BibitemOpen
  \bibfield  {author} {\bibinfo {author} {\bibfnamefont {Y.}~\bibnamefont {Kuramoto}},\ }\bibfield  {title} {\enquote {\bibinfo {title} {Diffusion-induced chaos in reaction systems},}\ }\href {\doibase 10.1143/PTPS.64.346} {\bibfield  {journal} {\bibinfo  {journal} {Progress of Theoretical Physics Supplement}\ }\textbf {\bibinfo {volume} {64}},\ \bibinfo {pages} {346--367} (\bibinfo {year} {1978})}\BibitemShut {NoStop}%
\bibitem [{\citenamefont {Sivashinsky}(1977)}]{sivashinsky1977nonlinear}%
  \BibitemOpen
  \bibfield  {author} {\bibinfo {author} {\bibfnamefont {G.~I.}\ \bibnamefont {Sivashinsky}},\ }\bibfield  {title} {\enquote {\bibinfo {title} {Nonlinear analysis of hydrodynamic instability in laminar flames---{I}. derivation of basic equations},}\ }\href {\doibase 10.1016/0094-5765(77)90096-0} {\bibfield  {journal} {\bibinfo  {journal} {Acta Astronautica}\ }\textbf {\bibinfo {volume} {4}},\ \bibinfo {pages} {1177--1206} (\bibinfo {year} {1977})}\BibitemShut {NoStop}%
\bibitem [{\citenamefont {Chirikov}(1979)}]{chirikov1979universal}%
  \BibitemOpen
  \bibfield  {author} {\bibinfo {author} {\bibfnamefont {B.~V.}\ \bibnamefont {Chirikov}},\ }\bibfield  {title} {\enquote {\bibinfo {title} {A universal instability of many-dimensional oscillator systems},}\ }\href {\doibase 10.1016/0370-1573(79)90023-1} {\bibfield  {journal} {\bibinfo  {journal} {Physics Reports}\ }\textbf {\bibinfo {volume} {52}},\ \bibinfo {pages} {263--379} (\bibinfo {year} {1979})}\BibitemShut {NoStop}%
\bibitem [{\citenamefont {Abdullaev}(2006)}]{abdullaev2006construction}%
  \BibitemOpen
  \bibfield  {author} {\bibinfo {author} {\bibfnamefont {S.~S.}\ \bibnamefont {Abdullaev}},\ }\href {\doibase 10.1007/b11554707} {\emph {\bibinfo {title} {Construction of Mappings for Hamiltonian Systems and Their Applications}}},\ \bibinfo {series} {Lecture Notes in Physics}, Vol.\ \bibinfo {volume} {691}\ (\bibinfo  {publisher} {Springer},\ \bibinfo {address} {Berlin},\ \bibinfo {year} {2006})\BibitemShut {NoStop}%
\bibitem [{\citenamefont {Hazeltine}\ and\ \citenamefont {Meiss}(2003)}]{hazeltine2003plasma}%
  \BibitemOpen
  \bibfield  {author} {\bibinfo {author} {\bibfnamefont {R.~D.}\ \bibnamefont {Hazeltine}}\ and\ \bibinfo {author} {\bibfnamefont {J.~D.}\ \bibnamefont {Meiss}},\ }\href@noop {} {\emph {\bibinfo {title} {Plasma Confinement}}}\ (\bibinfo  {publisher} {Dover Publications},\ \bibinfo {address} {Mineola, NY},\ \bibinfo {year} {2003})\BibitemShut {NoStop}%
\bibitem [{\citenamefont {Evans}\ \emph {et~al.}(2004)\citenamefont {Evans}, \citenamefont {Moyer}, \citenamefont {Thomas}, \citenamefont {Watkins}, \citenamefont {Osborne}, \citenamefont {Boedo}, \citenamefont {Doyle}, \citenamefont {Fenstermacher}, \citenamefont {Finken}, \citenamefont {Groebner}, \citenamefont {Groth}, \citenamefont {Harris}, \citenamefont {La~Haye}, \citenamefont {Lasnier}, \citenamefont {Masuzaki}, \citenamefont {Ohyabu}, \citenamefont {Pretty}, \citenamefont {Rhodes}, \citenamefont {Reimerdes}, \citenamefont {Rudakov}, \citenamefont {Schaffer}, \citenamefont {Wang},\ and\ \citenamefont {Zeng}}]{evans2004suppression}%
  \BibitemOpen
  \bibfield  {author} {\bibinfo {author} {\bibfnamefont {T.~E.}\ \bibnamefont {Evans}}, \bibinfo {author} {\bibfnamefont {R.~A.}\ \bibnamefont {Moyer}}, \bibinfo {author} {\bibfnamefont {P.~R.}\ \bibnamefont {Thomas}}, \bibinfo {author} {\bibfnamefont {J.~G.}\ \bibnamefont {Watkins}}, \bibinfo {author} {\bibfnamefont {T.~H.}\ \bibnamefont {Osborne}}, \bibinfo {author} {\bibfnamefont {J.~A.}\ \bibnamefont {Boedo}}, \bibinfo {author} {\bibfnamefont {E.~J.}\ \bibnamefont {Doyle}}, \bibinfo {author} {\bibfnamefont {M.~E.}\ \bibnamefont {Fenstermacher}}, \bibinfo {author} {\bibfnamefont {K.~H.}\ \bibnamefont {Finken}}, \bibinfo {author} {\bibfnamefont {R.~J.}\ \bibnamefont {Groebner}}, \bibinfo {author} {\bibfnamefont {M.}~\bibnamefont {Groth}}, \bibinfo {author} {\bibfnamefont {J.~H.}\ \bibnamefont {Harris}}, \bibinfo {author} {\bibfnamefont {R.~J.}\ \bibnamefont {La~Haye}}, \bibinfo {author} {\bibfnamefont {C.~J.}\ \bibnamefont {Lasnier}}, \bibinfo {author} {\bibfnamefont {S.}~\bibnamefont {Masuzaki}}, \bibinfo
  {author} {\bibfnamefont {N.}~\bibnamefont {Ohyabu}}, \bibinfo {author} {\bibfnamefont {D.~G.}\ \bibnamefont {Pretty}}, \bibinfo {author} {\bibfnamefont {T.~L.}\ \bibnamefont {Rhodes}}, \bibinfo {author} {\bibfnamefont {H.}~\bibnamefont {Reimerdes}}, \bibinfo {author} {\bibfnamefont {D.~L.}\ \bibnamefont {Rudakov}}, \bibinfo {author} {\bibfnamefont {M.~J.}\ \bibnamefont {Schaffer}}, \bibinfo {author} {\bibfnamefont {G.}~\bibnamefont {Wang}}, \ and\ \bibinfo {author} {\bibfnamefont {L.}~\bibnamefont {Zeng}},\ }\bibfield  {title} {\enquote {\bibinfo {title} {Suppression of large edge-localized modes in high-confinement {DIII-D} plasmas with a stochastic magnetic boundary},}\ }\href {\doibase 10.1103/PhysRevLett.92.235003} {\bibfield  {journal} {\bibinfo  {journal} {Physical Review Letters}\ }\textbf {\bibinfo {volume} {92}},\ \bibinfo {pages} {235003} (\bibinfo {year} {2004})}\BibitemShut {NoStop}%
\bibitem [{\citenamefont {Mitchell}\ \emph {et~al.}(2003{\natexlab{a}})\citenamefont {Mitchell}, \citenamefont {Handley}, \citenamefont {Tighe}, \citenamefont {Delos},\ and\ \citenamefont {Knudson}}]{Mitchell2003}%
  \BibitemOpen
  \bibfield  {author} {\bibinfo {author} {\bibfnamefont {K.~A.}\ \bibnamefont {Mitchell}}, \bibinfo {author} {\bibfnamefont {J.~P.}\ \bibnamefont {Handley}}, \bibinfo {author} {\bibfnamefont {B.}~\bibnamefont {Tighe}}, \bibinfo {author} {\bibfnamefont {J.~B.}\ \bibnamefont {Delos}}, \ and\ \bibinfo {author} {\bibfnamefont {S.~K.}\ \bibnamefont {Knudson}},\ }\bibfield  {title} {\enquote {\bibinfo {title} {Geometry and topology of escape. i. epistrophes},}\ }\href {\doibase 10.1063/1.1598311} {\bibfield  {journal} {\bibinfo  {journal} {Chaos: An Interdisciplinary Journal of Nonlinear Science}\ }\textbf {\bibinfo {volume} {13}},\ \bibinfo {pages} {880--891} (\bibinfo {year} {2003}{\natexlab{a}})}\BibitemShut {NoStop}%
\bibitem [{\citenamefont {Mitchell}\ \emph {et~al.}(2004)\citenamefont {Mitchell}, \citenamefont {Handley}, \citenamefont {Tighe}, \citenamefont {Flower},\ and\ \citenamefont {Delos}}]{Mitchell2004}%
  \BibitemOpen
  \bibfield  {author} {\bibinfo {author} {\bibfnamefont {K.~A.}\ \bibnamefont {Mitchell}}, \bibinfo {author} {\bibfnamefont {J.~P.}\ \bibnamefont {Handley}}, \bibinfo {author} {\bibfnamefont {B.}~\bibnamefont {Tighe}}, \bibinfo {author} {\bibfnamefont {A.}~\bibnamefont {Flower}}, \ and\ \bibinfo {author} {\bibfnamefont {J.~B.}\ \bibnamefont {Delos}},\ }\bibfield  {title} {\enquote {\bibinfo {title} {Analysis of chaos-induced pulse trains in the ionization of hydrogen},}\ }\href {\doibase 10.1103/PhysRevA.70.043407} {\bibfield  {journal} {\bibinfo  {journal} {Physical Review A}\ }\textbf {\bibinfo {volume} {70}},\ \bibinfo {pages} {043407} (\bibinfo {year} {2004})}\BibitemShut {NoStop}%
\bibitem [{\citenamefont {Jakubowski}\ \emph {et~al.}(2007)\citenamefont {Jakubowski}, \citenamefont {Wingen}, \citenamefont {Abdullaev}, \citenamefont {Finken}, \citenamefont {Lehnen}, \citenamefont {Spatschek},\ and\ \citenamefont {Wolf}}]{Jakubowski2007}%
  \BibitemOpen
  \bibfield  {author} {\bibinfo {author} {\bibfnamefont {M.~W.}\ \bibnamefont {Jakubowski}}, \bibinfo {author} {\bibfnamefont {A.}~\bibnamefont {Wingen}}, \bibinfo {author} {\bibfnamefont {S.~S.}\ \bibnamefont {Abdullaev}}, \bibinfo {author} {\bibfnamefont {K.~H.}\ \bibnamefont {Finken}}, \bibinfo {author} {\bibfnamefont {M.}~\bibnamefont {Lehnen}}, \bibinfo {author} {\bibfnamefont {K.~H.}\ \bibnamefont {Spatschek}}, \ and\ \bibinfo {author} {\bibfnamefont {R.~C.}\ \bibnamefont {Wolf}},\ }\bibfield  {title} {\enquote {\bibinfo {title} {Observation of the heteroclinic tangles in the heat flux pattern of the ergodic divertor at {TEXTOR}},}\ }\href {\doibase 10.1016/j.jnucmat.2007.01.158} {\bibfield  {journal} {\bibinfo  {journal} {Journal of Nuclear Materials}\ }\textbf {\bibinfo {volume} {363--365}},\ \bibinfo {pages} {371--376} (\bibinfo {year} {2007})}\BibitemShut {NoStop}%
\bibitem [{\citenamefont {Wingen}, \citenamefont {Evans},\ and\ \citenamefont {Spatschek}(2009{\natexlab{a}})}]{Wingen2009Footprints}%
  \BibitemOpen
  \bibfield  {author} {\bibinfo {author} {\bibfnamefont {A.}~\bibnamefont {Wingen}}, \bibinfo {author} {\bibfnamefont {T.~E.}\ \bibnamefont {Evans}}, \ and\ \bibinfo {author} {\bibfnamefont {K.~H.}\ \bibnamefont {Spatschek}},\ }\bibfield  {title} {\enquote {\bibinfo {title} {Footprint structures due to resonant magnetic perturbations in {DIII-D}},}\ }\href {\doibase 10.1063/1.3099053} {\bibfield  {journal} {\bibinfo  {journal} {Physics of Plasmas}\ }\textbf {\bibinfo {volume} {16}},\ \bibinfo {pages} {042504} (\bibinfo {year} {2009}{\natexlab{a}})}\BibitemShut {NoStop}%
\bibitem [{\citenamefont {Smiet}\ \emph {et~al.}(2025)\citenamefont {Smiet}, \citenamefont {Rais}, \citenamefont {Loizu},\ and\ \citenamefont {Davies}}]{Smiet2025}%
  \BibitemOpen
  \bibfield  {author} {\bibinfo {author} {\bibfnamefont {C.~B.}\ \bibnamefont {Smiet}}, \bibinfo {author} {\bibfnamefont {L.}~\bibnamefont {Rais}}, \bibinfo {author} {\bibfnamefont {J.}~\bibnamefont {Loizu}}, \ and\ \bibinfo {author} {\bibfnamefont {R.}~\bibnamefont {Davies}},\ }\bibfield  {title} {\enquote {\bibinfo {title} {Turnstile flux as a measure for chaotic transport in magnetic confinement fusion devices},}\ }\href {\doibase 10.1063/5.0275878} {\bibfield  {journal} {\bibinfo  {journal} {Chaos: An Interdisciplinary Journal of Nonlinear Science}\ }\textbf {\bibinfo {volume} {35}},\ \bibinfo {pages} {073129} (\bibinfo {year} {2025})}\BibitemShut {NoStop}%
\bibitem [{\citenamefont {Custodio}, \citenamefont {Sattari},\ and\ \citenamefont {Mitchell}(2025)}]{custodio2025hydrogen}%
  \BibitemOpen
  \bibfield  {author} {\bibinfo {author} {\bibfnamefont {E.~T.}\ \bibnamefont {Custodio}}, \bibinfo {author} {\bibfnamefont {S.}~\bibnamefont {Sattari}}, \ and\ \bibinfo {author} {\bibfnamefont {K.~A.}\ \bibnamefont {Mitchell}},\ }\bibfield  {title} {\enquote {\bibinfo {title} {{Computing classical escape rates from periodic orbits in chaotic hydrogen}},}\ }\href {\doibase 10.1063/5.0237613} {\bibfield  {journal} {\bibinfo  {journal} {Chaos: An Interdisciplinary Journal of Nonlinear Science}\ }\textbf {\bibinfo {volume} {35}},\ \bibinfo {pages} {053124} (\bibinfo {year} {2025})},\ \Eprint {http://arxiv.org/abs/2503.15710} {arXiv:2503.15710} \BibitemShut {NoStop}%
\bibitem [{\citenamefont {Goldston}\ and\ \citenamefont {Rutherford}(1995)}]{goldston1995introduction}%
  \BibitemOpen
  \bibfield  {author} {\bibinfo {author} {\bibfnamefont {R.~J.}\ \bibnamefont {Goldston}}\ and\ \bibinfo {author} {\bibfnamefont {P.~H.}\ \bibnamefont {Rutherford}},\ }\href {https://www.routledge.com/Introduction-to-Plasma-Physics/Goldston-Rutherford/p/book/9780750301831} {\emph {\bibinfo {title} {Introduction to Plasma Physics}}}\ (\bibinfo  {publisher} {Institute of Physics Publishing},\ \bibinfo {address} {Bristol},\ \bibinfo {year} {1995})\BibitemShut {NoStop}%
\bibitem [{\citenamefont {Cary}\ and\ \citenamefont {Littlejohn}(1983)}]{cary1983noncanonical}%
  \BibitemOpen
  \bibfield  {author} {\bibinfo {author} {\bibfnamefont {J.~R.}\ \bibnamefont {Cary}}\ and\ \bibinfo {author} {\bibfnamefont {R.~G.}\ \bibnamefont {Littlejohn}},\ }\bibfield  {title} {\enquote {\bibinfo {title} {Noncanonical hamiltonian mechanics and its application to magnetic field line flow},}\ }\href {\doibase 10.1016/0003-4916(83)90313-5} {\bibfield  {journal} {\bibinfo  {journal} {Annals of Physics}\ }\textbf {\bibinfo {volume} {151}},\ \bibinfo {pages} {1--34} (\bibinfo {year} {1983})}\BibitemShut {NoStop}%
\bibitem [{\citenamefont {Arnol'd}(1989)}]{arnold1989mathematical}%
  \BibitemOpen
  \bibfield  {author} {\bibinfo {author} {\bibfnamefont {V.~I.}\ \bibnamefont {Arnol'd}},\ }\href {\doibase 10.1007/978-1-4757-2063-1} {\emph {\bibinfo {title} {Mathematical Methods of Classical Mechanics}}},\ \bibinfo {edition} {2nd}\ ed.,\ \bibinfo {series} {Graduate Texts in Mathematics}, Vol.~\bibinfo {volume} {60}\ (\bibinfo  {publisher} {Springer},\ \bibinfo {address} {New York},\ \bibinfo {year} {1989})\BibitemShut {NoStop}%
\bibitem [{\citenamefont {Fitzpatrick}\ and\ \citenamefont {Hender}(1991)}]{fitzpatrick1991interaction}%
  \BibitemOpen
  \bibfield  {author} {\bibinfo {author} {\bibfnamefont {R.}~\bibnamefont {Fitzpatrick}}\ and\ \bibinfo {author} {\bibfnamefont {T.~C.}\ \bibnamefont {Hender}},\ }\bibfield  {title} {\enquote {\bibinfo {title} {The interaction of resonant magnetic perturbations with rotating plasmas},}\ }\href {\doibase 10.1063/1.859863} {\bibfield  {journal} {\bibinfo  {journal} {Physics of Fluids B: Plasma Physics}\ }\textbf {\bibinfo {volume} {3}},\ \bibinfo {pages} {644--673} (\bibinfo {year} {1991})}\BibitemShut {NoStop}%
\bibitem [{\citenamefont {Lichtenberg}\ and\ \citenamefont {Lieberman}(2013)}]{lichtenberg2013}%
  \BibitemOpen
  \bibfield  {author} {\bibinfo {author} {\bibfnamefont {A.~J.}\ \bibnamefont {Lichtenberg}}\ and\ \bibinfo {author} {\bibfnamefont {M.~A.}\ \bibnamefont {Lieberman}},\ }\href@noop {} {\emph {\bibinfo {title} {Regular and Chaotic Dynamics}}},\ Vol.~\bibinfo {volume} {38}\ (\bibinfo  {publisher} {Springer Science \& Business Media},\ \bibinfo {year} {2013})\BibitemShut {NoStop}%
\bibitem [{\citenamefont {Williams}, \citenamefont {Gudorf},\ and\ \citenamefont {Orlov}(2025)}]{Williams2025}%
  \BibitemOpen
  \bibfield  {author} {\bibinfo {author} {\bibfnamefont {S.~D.~V.}\ \bibnamefont {Williams}}, \bibinfo {author} {\bibfnamefont {M.~N.}\ \bibnamefont {Gudorf}}, \ and\ \bibinfo {author} {\bibfnamefont {D.~M.}\ \bibnamefont {Orlov}},\ }\bibfield  {title} {\enquote {\bibinfo {title} {Understanding plasma turbulence through exact coherent structures},}\ }\href {\doibase 10.1063/5.0282368} {\bibfield  {journal} {\bibinfo  {journal} {Physics of Plasmas}\ }\textbf {\bibinfo {volume} {32}},\ \bibinfo {pages} {092302} (\bibinfo {year} {2025})}\BibitemShut {NoStop}%
\bibitem [{\citenamefont {Evans}\ \emph {et~al.}(2007)\citenamefont {Evans}, \citenamefont {Joseph}, \citenamefont {Moyer}, \citenamefont {Fenstermacher}, \citenamefont {Lasnier},\ and\ \citenamefont {Yan}}]{evans2007separatrix}%
  \BibitemOpen
  \bibfield  {author} {\bibinfo {author} {\bibfnamefont {T.~E.}\ \bibnamefont {Evans}}, \bibinfo {author} {\bibfnamefont {I.}~\bibnamefont {Joseph}}, \bibinfo {author} {\bibfnamefont {R.~A.}\ \bibnamefont {Moyer}}, \bibinfo {author} {\bibfnamefont {M.~E.}\ \bibnamefont {Fenstermacher}}, \bibinfo {author} {\bibfnamefont {C.~J.}\ \bibnamefont {Lasnier}}, \ and\ \bibinfo {author} {\bibfnamefont {L.}~\bibnamefont {Yan}},\ }\bibfield  {title} {\enquote {\bibinfo {title} {Experimental and numerical studies of separatrix splitting and magnetic footprints in {DIII-D}},}\ }\href {\doibase 10.1016/j.jnucmat.2006.12.064} {\bibfield  {journal} {\bibinfo  {journal} {Journal of Nuclear Materials}\ }\textbf {\bibinfo {volume} {363--365}},\ \bibinfo {pages} {570--574} (\bibinfo {year} {2007})}\BibitemShut {NoStop}%
\bibitem [{\citenamefont {Luxon}(2002)}]{Luxon2002}%
  \BibitemOpen
  \bibfield  {author} {\bibinfo {author} {\bibfnamefont {J.~L.}\ \bibnamefont {Luxon}},\ }\bibfield  {title} {\enquote {\bibinfo {title} {A design retrospective of the {DIII-D} tokamak},}\ }\href {\doibase 10.1088/0029-5515/42/5/313} {\bibfield  {journal} {\bibinfo  {journal} {Nuclear Fusion}\ }\textbf {\bibinfo {volume} {42}},\ \bibinfo {pages} {614--633} (\bibinfo {year} {2002})}\BibitemShut {NoStop}%
\bibitem [{\citenamefont {Jackson}\ \emph {et~al.}(2003)\citenamefont {Jackson}, \citenamefont {Anderson}, \citenamefont {Bialek}, \citenamefont {Cary}, \citenamefont {Campbell}, \citenamefont {Garofalo}, \citenamefont {Hatcher}, \citenamefont {Kellman}, \citenamefont {La~Haye}, \citenamefont {Nagy}, \citenamefont {Navratil}, \citenamefont {Okabayashi}, \citenamefont {Pawley}, \citenamefont {Reimerdes}, \citenamefont {Scoville}, \citenamefont {Strait},\ and\ \citenamefont {Szymanski}}]{Jackson2003}%
  \BibitemOpen
  \bibfield  {author} {\bibinfo {author} {\bibfnamefont {G.~L.}\ \bibnamefont {Jackson}}, \bibinfo {author} {\bibfnamefont {P.~M.}\ \bibnamefont {Anderson}}, \bibinfo {author} {\bibfnamefont {J.}~\bibnamefont {Bialek}}, \bibinfo {author} {\bibfnamefont {W.~P.}\ \bibnamefont {Cary}}, \bibinfo {author} {\bibfnamefont {G.~L.}\ \bibnamefont {Campbell}}, \bibinfo {author} {\bibfnamefont {A.~M.}\ \bibnamefont {Garofalo}}, \bibinfo {author} {\bibfnamefont {R.}~\bibnamefont {Hatcher}}, \bibinfo {author} {\bibfnamefont {A.~G.}\ \bibnamefont {Kellman}}, \bibinfo {author} {\bibfnamefont {R.~J.}\ \bibnamefont {La~Haye}}, \bibinfo {author} {\bibfnamefont {A.}~\bibnamefont {Nagy}}, \bibinfo {author} {\bibfnamefont {G.~A.}\ \bibnamefont {Navratil}}, \bibinfo {author} {\bibfnamefont {M.}~\bibnamefont {Okabayashi}}, \bibinfo {author} {\bibfnamefont {C.~J.}\ \bibnamefont {Pawley}}, \bibinfo {author} {\bibfnamefont {H.}~\bibnamefont {Reimerdes}}, \bibinfo {author} {\bibfnamefont {J.~T.}\ \bibnamefont {Scoville}}, \bibinfo
  {author} {\bibfnamefont {E.~J.}\ \bibnamefont {Strait}}, \ and\ \bibinfo {author} {\bibfnamefont {D.~D.}\ \bibnamefont {Szymanski}},\ }\bibfield  {title} {\enquote {\bibinfo {title} {Initial results from the new internal magnetic field coils for resistive wall mode stabilization in the {DIII-D} tokamak},}\ }in\ \href@noop {} {\emph {\bibinfo {booktitle} {Proceedings of the 30th EPS Conference on Controlled Fusion and Plasma Physics}}},\ Vol.\ \bibinfo {volume} {27A}\ (\bibinfo {address} {St. Petersburg, Russia},\ \bibinfo {year} {2003})\ pp.\ \bibinfo {pages} {P--4.47}\BibitemShut {NoStop}%
\bibitem [{\citenamefont {Schmitz}\ \emph {et~al.}(2019)\citenamefont {Schmitz}, \citenamefont {Kriete}, \citenamefont {Wilcox}, \citenamefont {Rhodes}, \citenamefont {Zeng}, \citenamefont {Yan}, \citenamefont {McKee}, \citenamefont {Evans}, \citenamefont {Paz-Soldan}, \citenamefont {Gohil}, \citenamefont {Lyons}, \citenamefont {Petty}, \citenamefont {Orlov},\ and\ \citenamefont {Marinoni}}]{schmitz2019lh}%
  \BibitemOpen
  \bibfield  {author} {\bibinfo {author} {\bibfnamefont {L.}~\bibnamefont {Schmitz}}, \bibinfo {author} {\bibfnamefont {D.~M.}\ \bibnamefont {Kriete}}, \bibinfo {author} {\bibfnamefont {R.~S.}\ \bibnamefont {Wilcox}}, \bibinfo {author} {\bibfnamefont {T.~L.}\ \bibnamefont {Rhodes}}, \bibinfo {author} {\bibfnamefont {L.}~\bibnamefont {Zeng}}, \bibinfo {author} {\bibfnamefont {Z.}~\bibnamefont {Yan}}, \bibinfo {author} {\bibfnamefont {G.~R.}\ \bibnamefont {McKee}}, \bibinfo {author} {\bibfnamefont {T.~E.}\ \bibnamefont {Evans}}, \bibinfo {author} {\bibfnamefont {C.}~\bibnamefont {Paz-Soldan}}, \bibinfo {author} {\bibfnamefont {P.}~\bibnamefont {Gohil}}, \bibinfo {author} {\bibfnamefont {B.}~\bibnamefont {Lyons}}, \bibinfo {author} {\bibfnamefont {C.~C.}\ \bibnamefont {Petty}}, \bibinfo {author} {\bibfnamefont {D.}~\bibnamefont {Orlov}}, \ and\ \bibinfo {author} {\bibfnamefont {A.}~\bibnamefont {Marinoni}},\ }\bibfield  {title} {\enquote {\bibinfo {title} {L--h transition trigger physics in iter-similar plasmas
  with applied {$n=3$} magnetic perturbations},}\ }\href {\doibase 10.1088/1741-4326/ab36bf} {\bibfield  {journal} {\bibinfo  {journal} {Nuclear Fusion}\ }\textbf {\bibinfo {volume} {59}},\ \bibinfo {pages} {126010} (\bibinfo {year} {2019})}\BibitemShut {NoStop}%
\bibitem [{\citenamefont {Wingen}, \citenamefont {Evans},\ and\ \citenamefont {Spatschek}(2009{\natexlab{b}})}]{wingen2009highresolution}%
  \BibitemOpen
  \bibfield  {author} {\bibinfo {author} {\bibfnamefont {A.}~\bibnamefont {Wingen}}, \bibinfo {author} {\bibfnamefont {T.~E.}\ \bibnamefont {Evans}}, \ and\ \bibinfo {author} {\bibfnamefont {K.~H.}\ \bibnamefont {Spatschek}},\ }\bibfield  {title} {\enquote {\bibinfo {title} {High-resolution numerical studies of separatrix splitting due to non-axisymmetric perturbation in {DIII-D}},}\ }\href {\doibase 10.1088/0029-5515/49/5/055027} {\bibfield  {journal} {\bibinfo  {journal} {Nuclear Fusion}\ }\textbf {\bibinfo {volume} {49}},\ \bibinfo {pages} {055027} (\bibinfo {year} {2009}{\natexlab{b}})}\BibitemShut {NoStop}%
\bibitem [{\citenamefont {Kostadinova}\ \emph {et~al.}(2023)\citenamefont {Kostadinova}, \citenamefont {Orlov}, \citenamefont {Koepke}, \citenamefont {Skiff},\ and\ \citenamefont {Austin}}]{Kostadinova2023}%
  \BibitemOpen
  \bibfield  {author} {\bibinfo {author} {\bibfnamefont {E.~G.}\ \bibnamefont {Kostadinova}}, \bibinfo {author} {\bibfnamefont {D.~M.}\ \bibnamefont {Orlov}}, \bibinfo {author} {\bibfnamefont {M.}~\bibnamefont {Koepke}}, \bibinfo {author} {\bibfnamefont {F.}~\bibnamefont {Skiff}}, \ and\ \bibinfo {author} {\bibfnamefont {M.~E.}\ \bibnamefont {Austin}},\ }\bibfield  {title} {\enquote {\bibinfo {title} {Energetic electron transport in magnetic fields with island chains and stochastic regions},}\ }\href {\doibase 10.1017/S0022377823000879} {\bibfield  {journal} {\bibinfo  {journal} {Journal of Plasma Physics}\ }\textbf {\bibinfo {volume} {89}},\ \bibinfo {pages} {905890420} (\bibinfo {year} {2023})}\BibitemShut {NoStop}%
\bibitem [{\citenamefont {Eskew}\ \emph {et~al.}(2026)\citenamefont {Eskew}, \citenamefont {Orlov}, \citenamefont {Andrew}, \citenamefont {Bursch}, \citenamefont {Koepke}, \citenamefont {Skiff}, \citenamefont {Austin}, \citenamefont {Cote}, \citenamefont {Marini},\ and\ \citenamefont {Kostadinova}}]{Eskew2026}%
  \BibitemOpen
  \bibfield  {author} {\bibinfo {author} {\bibfnamefont {J.}~\bibnamefont {Eskew}}, \bibinfo {author} {\bibfnamefont {D.~M.}\ \bibnamefont {Orlov}}, \bibinfo {author} {\bibfnamefont {B.}~\bibnamefont {Andrew}}, \bibinfo {author} {\bibfnamefont {E.}~\bibnamefont {Bursch}}, \bibinfo {author} {\bibfnamefont {M.}~\bibnamefont {Koepke}}, \bibinfo {author} {\bibfnamefont {F.}~\bibnamefont {Skiff}}, \bibinfo {author} {\bibfnamefont {M.~E.}\ \bibnamefont {Austin}}, \bibinfo {author} {\bibfnamefont {T.}~\bibnamefont {Cote}}, \bibinfo {author} {\bibfnamefont {C.}~\bibnamefont {Marini}}, \ and\ \bibinfo {author} {\bibfnamefont {E.~G.}\ \bibnamefont {Kostadinova}},\ }\bibfield  {title} {\enquote {\bibinfo {title} {Effect of controlled magnetic island bifurcation on electron diffusion},}\ }\href {\doibase 10.1088/1741-4326/ae6ab3} {\bibfield  {journal} {\bibinfo  {journal} {Nuclear Fusion}\ }\textbf {\bibinfo {volume} {66}},\ \bibinfo {pages} {066022} (\bibinfo {year} {2026})}\BibitemShut {NoStop}%
\bibitem [{\citenamefont {Khavin}, \citenamefont {Dudkovskaia},\ and\ \citenamefont {Orlov}(2026)}]{Khavin2026}%
  \BibitemOpen
  \bibfield  {author} {\bibinfo {author} {\bibfnamefont {V.}~\bibnamefont {Khavin}}, \bibinfo {author} {\bibfnamefont {A.}~\bibnamefont {Dudkovskaia}}, \ and\ \bibinfo {author} {\bibfnamefont {D.~M.}\ \bibnamefont {Orlov}},\ }\href@noop {} {\enquote {\bibinfo {title} {Chaotic separatrix layer effects on drift kinetic solutions near magnetic island},}\ } (\bibinfo {year} {2026}),\ \bibinfo {note} {manuscript in preparation}\BibitemShut {NoStop}%
\bibitem [{\citenamefont {Sattari}\ and\ \citenamefont {Mitchell}(2017)}]{sattari2017periodic}%
  \BibitemOpen
  \bibfield  {author} {\bibinfo {author} {\bibfnamefont {S.}~\bibnamefont {Sattari}}\ and\ \bibinfo {author} {\bibfnamefont {K.~A.}\ \bibnamefont {Mitchell}},\ }\bibfield  {title} {\enquote {\bibinfo {title} {Using periodic orbits to compute chaotic transport rates between resonance zones},}\ }\href {\doibase 10.1063/1.4998219} {\bibfield  {journal} {\bibinfo  {journal} {Chaos: An Interdisciplinary Journal of Nonlinear Science}\ }\textbf {\bibinfo {volume} {27}},\ \bibinfo {pages} {113104} (\bibinfo {year} {2017})}\BibitemShut {NoStop}%
\bibitem [{\citenamefont {Ferraro}(2012)}]{ferraro2012twofluid}%
  \BibitemOpen
  \bibfield  {author} {\bibinfo {author} {\bibfnamefont {N.~M.}\ \bibnamefont {Ferraro}},\ }\bibfield  {title} {\enquote {\bibinfo {title} {Calculations of two-fluid linear response to non-axisymmetric fields in tokamaks},}\ }\href {\doibase 10.1063/1.3694657} {\bibfield  {journal} {\bibinfo  {journal} {Physics of Plasmas}\ }\textbf {\bibinfo {volume} {19}},\ \bibinfo {pages} {056105} (\bibinfo {year} {2012})}\BibitemShut {NoStop}%
\bibitem [{\citenamefont {Fitzpatrick}(1998)}]{fitzpatrick1998bifurcated}%
  \BibitemOpen
  \bibfield  {author} {\bibinfo {author} {\bibfnamefont {R.}~\bibnamefont {Fitzpatrick}},\ }\bibfield  {title} {\enquote {\bibinfo {title} {Bifurcated states of a rotating tokamak plasma in the presence of a static error-field},}\ }\href {\doibase 10.1063/1.873000} {\bibfield  {journal} {\bibinfo  {journal} {Physics of Plasmas}\ }\textbf {\bibinfo {volume} {5}},\ \bibinfo {pages} {3325--3341} (\bibinfo {year} {1998})}\BibitemShut {NoStop}%
\bibitem [{\citenamefont {Kim}\ \emph {et~al.}(2015)\citenamefont {Kim}, \citenamefont {Ahn}, \citenamefont {Scotti}, \citenamefont {Park},\ and\ \citenamefont {Menard}}]{kim2015ideal}%
  \BibitemOpen
  \bibfield  {author} {\bibinfo {author} {\bibfnamefont {K.}~\bibnamefont {Kim}}, \bibinfo {author} {\bibfnamefont {J.-W.}\ \bibnamefont {Ahn}}, \bibinfo {author} {\bibfnamefont {F.}~\bibnamefont {Scotti}}, \bibinfo {author} {\bibfnamefont {J.-K.}\ \bibnamefont {Park}}, \ and\ \bibinfo {author} {\bibfnamefont {J.~E.}\ \bibnamefont {Menard}},\ }\bibfield  {title} {\enquote {\bibinfo {title} {Ideal plasma response to vacuum magnetic fields with resonant magnetic perturbations in non-axisymmetric tokamaks},}\ }\href {\doibase 10.1088/0741-3335/57/10/104002} {\bibfield  {journal} {\bibinfo  {journal} {Plasma Physics and Controlled Fusion}\ }\textbf {\bibinfo {volume} {57}},\ \bibinfo {pages} {104002} (\bibinfo {year} {2015})}\BibitemShut {NoStop}%
\bibitem [{\citenamefont {MacKay}, \citenamefont {Meiss},\ and\ \citenamefont {Percival}(1987)}]{mackay1987resonances}%
  \BibitemOpen
  \bibfield  {author} {\bibinfo {author} {\bibfnamefont {R.~S.}\ \bibnamefont {MacKay}}, \bibinfo {author} {\bibfnamefont {J.~D.}\ \bibnamefont {Meiss}}, \ and\ \bibinfo {author} {\bibfnamefont {I.~C.}\ \bibnamefont {Percival}},\ }\bibfield  {title} {\enquote {\bibinfo {title} {{Resonances in area preserving maps}},}\ }\href {\doibase 10.1016/0167-2789(87)90002-9} {\bibfield  {journal} {\bibinfo  {journal} {Physica D: Nonlinear Phenomena}\ }\textbf {\bibinfo {volume} {27}},\ \bibinfo {pages} {1--20} (\bibinfo {year} {1987})}\BibitemShut {NoStop}%
\bibitem [{\citenamefont {Wiggins}(1990)}]{wiggins1990geometry}%
  \BibitemOpen
  \bibfield  {author} {\bibinfo {author} {\bibfnamefont {S.}~\bibnamefont {Wiggins}},\ }\bibfield  {title} {\enquote {\bibinfo {title} {{On the geometry of transport in phase space I. Transport in $k$-degree-of-freedom {H}amiltonian systems, $2 \leq k < \infty$}},}\ }\href {\doibase 10.1016/0167-2789(90)90159-M} {\bibfield  {journal} {\bibinfo  {journal} {Physica D: Nonlinear Phenomena}\ }\textbf {\bibinfo {volume} {44}},\ \bibinfo {pages} {471--501} (\bibinfo {year} {1990})}\BibitemShut {NoStop}%
\bibitem [{\citenamefont {Rom-Kedar}(1994)}]{romkedar1994homoclinic}%
  \BibitemOpen
  \bibfield  {author} {\bibinfo {author} {\bibfnamefont {V.}~\bibnamefont {Rom-Kedar}},\ }\bibfield  {title} {\enquote {\bibinfo {title} {{Homoclinic tangles---classification and applications}},}\ }\href {\doibase 10.1088/0951-7715/7/2/008} {\bibfield  {journal} {\bibinfo  {journal} {Nonlinearity}\ }\textbf {\bibinfo {volume} {7}},\ \bibinfo {pages} {441--473} (\bibinfo {year} {1994})}\BibitemShut {NoStop}%
\bibitem [{\citenamefont {Mitchell}(2012)}]{mitchell2012partitioning}%
  \BibitemOpen
  \bibfield  {author} {\bibinfo {author} {\bibfnamefont {K.~A.}\ \bibnamefont {Mitchell}},\ }\bibfield  {title} {\enquote {\bibinfo {title} {{Partitioning two-dimensional mixed phase spaces}},}\ }\href {\doibase 10.1016/j.physd.2012.07.004} {\bibfield  {journal} {\bibinfo  {journal} {Physica D: Nonlinear Phenomena}\ }\textbf {\bibinfo {volume} {241}},\ \bibinfo {pages} {1718--1734} (\bibinfo {year} {2012})}\BibitemShut {NoStop}%
\bibitem [{\citenamefont {Lomel\'{i}}\ and\ \citenamefont {Meiss}(2000)}]{lomeli2000heteroclinic}%
  \BibitemOpen
  \bibfield  {author} {\bibinfo {author} {\bibfnamefont {H.~E.}\ \bibnamefont {Lomel\'{i}}}\ and\ \bibinfo {author} {\bibfnamefont {J.~D.}\ \bibnamefont {Meiss}},\ }\bibfield  {title} {\enquote {\bibinfo {title} {{Heteroclinic primary intersections and codimension one Melnikov method for volume-preserving maps}},}\ }\href {\doibase 10.1063/1.166480} {\bibfield  {journal} {\bibinfo  {journal} {Chaos: An Interdisciplinary Journal of Nonlinear Science}\ }\textbf {\bibinfo {volume} {10}},\ \bibinfo {pages} {109--121} (\bibinfo {year} {2000})}\BibitemShut {NoStop}%
\bibitem [{\citenamefont {Lomel\'{i}}\ and\ \citenamefont {Meiss}(2009)}]{lomeli2009resonance}%
  \BibitemOpen
  \bibfield  {author} {\bibinfo {author} {\bibfnamefont {H.~E.}\ \bibnamefont {Lomel\'{i}}}\ and\ \bibinfo {author} {\bibfnamefont {J.~D.}\ \bibnamefont {Meiss}},\ }\bibfield  {title} {\enquote {\bibinfo {title} {{Resonance zones and lobe volumes for exact volume-preserving maps}},}\ }\href {\doibase 10.1088/0951-7715/22/8/001} {\bibfield  {journal} {\bibinfo  {journal} {Nonlinearity}\ }\textbf {\bibinfo {volume} {22}},\ \bibinfo {pages} {1761--1789} (\bibinfo {year} {2009})}\BibitemShut {NoStop}%
\bibitem [{\citenamefont {Mitchell}\ \emph {et~al.}(2003{\natexlab{b}})\citenamefont {Mitchell}, \citenamefont {Handley}, \citenamefont {Tighe}, \citenamefont {Delos},\ and\ \citenamefont {Knudson}}]{mitchell2003escape1}%
  \BibitemOpen
  \bibfield  {author} {\bibinfo {author} {\bibfnamefont {K.~A.}\ \bibnamefont {Mitchell}}, \bibinfo {author} {\bibfnamefont {J.~P.}\ \bibnamefont {Handley}}, \bibinfo {author} {\bibfnamefont {B.}~\bibnamefont {Tighe}}, \bibinfo {author} {\bibfnamefont {J.~B.}\ \bibnamefont {Delos}}, \ and\ \bibinfo {author} {\bibfnamefont {S.~K.}\ \bibnamefont {Knudson}},\ }\bibfield  {title} {\enquote {\bibinfo {title} {{Geometry and topology of escape. I. Epistrophes}},}\ }\href {\doibase 10.1063/1.1598311} {\bibfield  {journal} {\bibinfo  {journal} {Chaos: An Interdisciplinary Journal of Nonlinear Science}\ }\textbf {\bibinfo {volume} {13}},\ \bibinfo {pages} {880--891} (\bibinfo {year} {2003}{\natexlab{b}})}\BibitemShut {NoStop}%
\bibitem [{\citenamefont {Mitchell}\ \emph {et~al.}(2003{\natexlab{c}})\citenamefont {Mitchell}, \citenamefont {Handley}, \citenamefont {Delos},\ and\ \citenamefont {Knudson}}]{mitchell2003escape2}%
  \BibitemOpen
  \bibfield  {author} {\bibinfo {author} {\bibfnamefont {K.~A.}\ \bibnamefont {Mitchell}}, \bibinfo {author} {\bibfnamefont {J.~P.}\ \bibnamefont {Handley}}, \bibinfo {author} {\bibfnamefont {J.~B.}\ \bibnamefont {Delos}}, \ and\ \bibinfo {author} {\bibfnamefont {S.~K.}\ \bibnamefont {Knudson}},\ }\bibfield  {title} {\enquote {\bibinfo {title} {{Geometry and topology of escape. II. Homotopic lobe dynamics}},}\ }\href {\doibase 10.1063/1.1598312} {\bibfield  {journal} {\bibinfo  {journal} {Chaos: An Interdisciplinary Journal of Nonlinear Science}\ }\textbf {\bibinfo {volume} {13}},\ \bibinfo {pages} {892--902} (\bibinfo {year} {2003}{\natexlab{c}})},\ \Eprint {http://arxiv.org/abs/nlin/0307020} {arXiv:nlin/0307020} \BibitemShut {NoStop}%
\bibitem [{\citenamefont {Cvitanovi{\'c}}\ \emph {et~al.}(2025)\citenamefont {Cvitanovi{\'c}}, \citenamefont {Artuso}, \citenamefont {Mainieri}, \citenamefont {Tanner},\ and\ \citenamefont {Vattay}}]{cvitanovic2025chaosbook}%
  \BibitemOpen
  \bibfield  {author} {\bibinfo {author} {\bibfnamefont {P.}~\bibnamefont {Cvitanovi{\'c}}}, \bibinfo {author} {\bibfnamefont {R.}~\bibnamefont {Artuso}}, \bibinfo {author} {\bibfnamefont {R.}~\bibnamefont {Mainieri}}, \bibinfo {author} {\bibfnamefont {G.}~\bibnamefont {Tanner}}, \ and\ \bibinfo {author} {\bibfnamefont {G.}~\bibnamefont {Vattay}},\ }\href {https://chaosbook.org/chapters/ChaosBook.pdf} {\emph {\bibinfo {title} {Chaos: Classical and Quantum}}},\ \bibinfo {edition} {17th}\ ed.\ (\bibinfo  {publisher} {ChaosBook.org},\ \bibinfo {year} {2025})\ \bibinfo {note} {open webbook}\BibitemShut {NoStop}%
\bibitem [{\citenamefont {Wiggins}(1992)}]{wiggins1992chaotic}%
  \BibitemOpen
  \bibfield  {author} {\bibinfo {author} {\bibfnamefont {S.}~\bibnamefont {Wiggins}},\ }\href {\doibase 10.1007/978-1-4757-3896-4} {\emph {\bibinfo {title} {Chaotic Transport in Dynamical Systems}}},\ \bibinfo {series} {Interdisciplinary Applied Mathematics}, Vol.~\bibinfo {volume} {2}\ (\bibinfo  {publisher} {Springer},\ \bibinfo {address} {New York},\ \bibinfo {year} {1992})\BibitemShut {NoStop}%
\bibitem [{\citenamefont {Rechester}\ and\ \citenamefont {Rosenbluth}(1978)}]{rechester1978electron}%
  \BibitemOpen
  \bibfield  {author} {\bibinfo {author} {\bibfnamefont {A.~B.}\ \bibnamefont {Rechester}}\ and\ \bibinfo {author} {\bibfnamefont {M.~N.}\ \bibnamefont {Rosenbluth}},\ }\bibfield  {title} {\enquote {\bibinfo {title} {Electron heat transport in a tokamak with destroyed magnetic surfaces},}\ }\href {\doibase 10.1103/PhysRevLett.40.38} {\bibfield  {journal} {\bibinfo  {journal} {Physical Review Letters}\ }\textbf {\bibinfo {volume} {40}},\ \bibinfo {pages} {38--41} (\bibinfo {year} {1978})}\BibitemShut {NoStop}%
\bibitem [{\citenamefont {Alciatore}\ and\ \citenamefont {Miranda}(1995)}]{alciatore1995winding}%
  \BibitemOpen
  \bibfield  {author} {\bibinfo {author} {\bibfnamefont {D.~G.}\ \bibnamefont {Alciatore}}\ and\ \bibinfo {author} {\bibfnamefont {R.}~\bibnamefont {Miranda}},\ }\href {https://app1.icgc.cat/bd/mapes/mapcataleg/movido_de_php/etc/point_in_polygon.pdf} {\enquote {\bibinfo {title} {A winding number and point-in-polygon algorithm},}\ }\bibinfo {type} {Research Report}\ (\bibinfo  {institution} {Colorado State University},\ \bibinfo {address} {Fort Collins, Colorado},\ \bibinfo {year} {1995})\BibitemShut {NoStop}%
\bibitem [{\citenamefont {Hormann}\ and\ \citenamefont {Agathos}(2001)}]{hormann2001point}%
  \BibitemOpen
  \bibfield  {author} {\bibinfo {author} {\bibfnamefont {K.}~\bibnamefont {Hormann}}\ and\ \bibinfo {author} {\bibfnamefont {A.}~\bibnamefont {Agathos}},\ }\bibfield  {title} {\enquote {\bibinfo {title} {The point in polygon problem for arbitrary polygons},}\ }\href {\doibase 10.1016/S0925-7721(01)00012-8} {\bibfield  {journal} {\bibinfo  {journal} {Computational Geometry}\ }\textbf {\bibinfo {volume} {20}},\ \bibinfo {pages} {131--144} (\bibinfo {year} {2001})}\BibitemShut {NoStop}%
\bibitem [{\citenamefont {MacKay}, \citenamefont {Meiss},\ and\ \citenamefont {Percival}(1984)}]{mackay1984transport}%
  \BibitemOpen
  \bibfield  {author} {\bibinfo {author} {\bibfnamefont {R.~S.}\ \bibnamefont {MacKay}}, \bibinfo {author} {\bibfnamefont {J.~D.}\ \bibnamefont {Meiss}}, \ and\ \bibinfo {author} {\bibfnamefont {I.~C.}\ \bibnamefont {Percival}},\ }\bibfield  {title} {\enquote {\bibinfo {title} {Transport in hamiltonian systems},}\ }\href {\doibase 10.1016/0167-2789(84)90270-7} {\bibfield  {journal} {\bibinfo  {journal} {Physica D: Nonlinear Phenomena}\ }\textbf {\bibinfo {volume} {13}},\ \bibinfo {pages} {55--81} (\bibinfo {year} {1984})}\BibitemShut {NoStop}%
\bibitem [{\citenamefont {Altmann}, \citenamefont {Portela},\ and\ \citenamefont {T{\'e}l}(2013)}]{altmann2013leaking}%
  \BibitemOpen
  \bibfield  {author} {\bibinfo {author} {\bibfnamefont {E.~G.}\ \bibnamefont {Altmann}}, \bibinfo {author} {\bibfnamefont {J.~S.~E.}\ \bibnamefont {Portela}}, \ and\ \bibinfo {author} {\bibfnamefont {T.}~\bibnamefont {T{\'e}l}},\ }\bibfield  {title} {\enquote {\bibinfo {title} {Leaking chaotic systems},}\ }\href {\doibase 10.1103/RevModPhys.85.869} {\bibfield  {journal} {\bibinfo  {journal} {Reviews of Modern Physics}\ }\textbf {\bibinfo {volume} {85}},\ \bibinfo {pages} {869--918} (\bibinfo {year} {2013})}\BibitemShut {NoStop}%
\bibitem [{\citenamefont {Mitchell}(2009)}]{mitchell2009topology}%
  \BibitemOpen
  \bibfield  {author} {\bibinfo {author} {\bibfnamefont {K.~A.}\ \bibnamefont {Mitchell}},\ }\bibfield  {title} {\enquote {\bibinfo {title} {The topology of nested homoclinic and heteroclinic tangles},}\ }\href {\doibase 10.1016/j.physd.2009.01.004} {\bibfield  {journal} {\bibinfo  {journal} {Physica D: Nonlinear Phenomena}\ }\textbf {\bibinfo {volume} {238}},\ \bibinfo {pages} {737--763} (\bibinfo {year} {2009})}\BibitemShut {NoStop}%
\bibitem [{\citenamefont {Naik}, \citenamefont {Lekien},\ and\ \citenamefont {Ross}(2017)}]{naik2017computational}%
  \BibitemOpen
  \bibfield  {author} {\bibinfo {author} {\bibfnamefont {S.}~\bibnamefont {Naik}}, \bibinfo {author} {\bibfnamefont {F.}~\bibnamefont {Lekien}}, \ and\ \bibinfo {author} {\bibfnamefont {S.~D.}\ \bibnamefont {Ross}},\ }\bibfield  {title} {\enquote {\bibinfo {title} {Computational method for phase space transport with applications to lobe dynamics and rate of escape},}\ }\href {\doibase 10.1134/S1560354717030078} {\bibfield  {journal} {\bibinfo  {journal} {Regular and Chaotic Dynamics}\ }\textbf {\bibinfo {volume} {22}},\ \bibinfo {pages} {272--297} (\bibinfo {year} {2017})}\BibitemShut {NoStop}%
\bibitem [{\citenamefont {Gavrilova}\ and\ \citenamefont {Rokne}(2000)}]{gavrilova2000reliable}%
  \BibitemOpen
  \bibfield  {author} {\bibinfo {author} {\bibfnamefont {M.~L.}\ \bibnamefont {Gavrilova}}\ and\ \bibinfo {author} {\bibfnamefont {J.~G.}\ \bibnamefont {Rokne}},\ }\bibfield  {title} {\enquote {\bibinfo {title} {Reliable line segment intersection testing},}\ }\href {\doibase 10.1016/S0010-4485(00)00050-6} {\bibfield  {journal} {\bibinfo  {journal} {Computer-Aided Design}\ }\textbf {\bibinfo {volume} {32}},\ \bibinfo {pages} {737--745} (\bibinfo {year} {2000})}\BibitemShut {NoStop}%
\bibitem [{\citenamefont {Deshmukh}\ \emph {et~al.}(2021)\citenamefont {Deshmukh}, \citenamefont {Bradley}, \citenamefont {Garland},\ and\ \citenamefont {Meiss}}]{deshmukh2021scaling}%
  \BibitemOpen
  \bibfield  {author} {\bibinfo {author} {\bibfnamefont {V.}~\bibnamefont {Deshmukh}}, \bibinfo {author} {\bibfnamefont {E.}~\bibnamefont {Bradley}}, \bibinfo {author} {\bibfnamefont {J.}~\bibnamefont {Garland}}, \ and\ \bibinfo {author} {\bibfnamefont {J.~D.}\ \bibnamefont {Meiss}},\ }\bibfield  {title} {\enquote {\bibinfo {title} {Toward automated extraction and characterization of scaling regions in dynamical systems},}\ }\href {\doibase 10.1063/5.0069365} {\bibfield  {journal} {\bibinfo  {journal} {Chaos: An Interdisciplinary Journal of Nonlinear Science}\ }\textbf {\bibinfo {volume} {31}},\ \bibinfo {pages} {123102} (\bibinfo {year} {2021})}\BibitemShut {NoStop}%
\bibitem [{\citenamefont {Schwarz}(1978)}]{schwarz1978estimating}%
  \BibitemOpen
  \bibfield  {author} {\bibinfo {author} {\bibfnamefont {G.}~\bibnamefont {Schwarz}},\ }\bibfield  {title} {\enquote {\bibinfo {title} {Estimating the dimension of a model},}\ }\href {\doibase 10.1214/aos/1176344136} {\bibfield  {journal} {\bibinfo  {journal} {The Annals of Statistics}\ }\textbf {\bibinfo {volume} {6}},\ \bibinfo {pages} {461--464} (\bibinfo {year} {1978})}\BibitemShut {NoStop}%
\bibitem [{\citenamefont {Katz}(1981)}]{katz1981criteria}%
  \BibitemOpen
  \bibfield  {author} {\bibinfo {author} {\bibfnamefont {R.~W.}\ \bibnamefont {Katz}},\ }\bibfield  {title} {\enquote {\bibinfo {title} {On some criteria for estimating the order of a markov chain},}\ }\href {\doibase 10.1080/00401706.1981.10486293} {\bibfield  {journal} {\bibinfo  {journal} {Technometrics}\ }\textbf {\bibinfo {volume} {23}},\ \bibinfo {pages} {243--249} (\bibinfo {year} {1981})}\BibitemShut {NoStop}%
\bibitem [{\citenamefont {Kass}\ and\ \citenamefont {Raftery}(1995)}]{kass1995bayes}%
  \BibitemOpen
  \bibfield  {author} {\bibinfo {author} {\bibfnamefont {R.~E.}\ \bibnamefont {Kass}}\ and\ \bibinfo {author} {\bibfnamefont {A.~E.}\ \bibnamefont {Raftery}},\ }\bibfield  {title} {\enquote {\bibinfo {title} {Bayes factors},}\ }\href {\doibase 10.1080/01621459.1995.10476572} {\bibfield  {journal} {\bibinfo  {journal} {Journal of the American Statistical Association}\ }\textbf {\bibinfo {volume} {90}},\ \bibinfo {pages} {773--795} (\bibinfo {year} {1995})}\BibitemShut {NoStop}%
\bibitem [{\citenamefont {Clopper}\ and\ \citenamefont {Pearson}(1934)}]{clopper1934use}%
  \BibitemOpen
  \bibfield  {author} {\bibinfo {author} {\bibfnamefont {C.~J.}\ \bibnamefont {Clopper}}\ and\ \bibinfo {author} {\bibfnamefont {E.~S.}\ \bibnamefont {Pearson}},\ }\bibfield  {title} {\enquote {\bibinfo {title} {The use of confidence or fiducial limits illustrated in the case of the binomial},}\ }\href {\doibase 10.1093/biomet/26.4.404} {\bibfield  {journal} {\bibinfo  {journal} {Biometrika}\ }\textbf {\bibinfo {volume} {26}},\ \bibinfo {pages} {404--413} (\bibinfo {year} {1934})}\BibitemShut {NoStop}%
\bibitem [{\citenamefont {Cameron}\ and\ \citenamefont {Trivedi}(1986)}]{cameron1986econometric}%
  \BibitemOpen
  \bibfield  {author} {\bibinfo {author} {\bibfnamefont {A.~C.}\ \bibnamefont {Cameron}}\ and\ \bibinfo {author} {\bibfnamefont {P.~K.}\ \bibnamefont {Trivedi}},\ }\bibfield  {title} {\enquote {\bibinfo {title} {Econometric models based on count data: Comparisons and applications of some estimators and tests},}\ }\href {\doibase 10.1002/jae.3950010104} {\bibfield  {journal} {\bibinfo  {journal} {Journal of Applied Econometrics}\ }\textbf {\bibinfo {volume} {1}},\ \bibinfo {pages} {29--53} (\bibinfo {year} {1986})}\BibitemShut {NoStop}%
\bibitem [{\citenamefont {Efron}(1979)}]{efron1979bootstrap}%
  \BibitemOpen
  \bibfield  {author} {\bibinfo {author} {\bibfnamefont {B.}~\bibnamefont {Efron}},\ }\bibfield  {title} {\enquote {\bibinfo {title} {Bootstrap methods: Another look at the jackknife},}\ }\href {\doibase 10.1214/aos/1176344552} {\bibfield  {journal} {\bibinfo  {journal} {The Annals of Statistics}\ }\textbf {\bibinfo {volume} {7}},\ \bibinfo {pages} {1--26} (\bibinfo {year} {1979})}\BibitemShut {NoStop}%
\end{thebibliography}
\end{document}